\documentclass[preprint]{aastex}
\usepackage{multirow}
\usepackage{graphicx}
\usepackage{color}
\usepackage{rotating}

\shortauthors{Wang}

\begin{document}
\title{How Close Are We To Detecting Earth-like Planets in the Habitable Zone Using the Radial Velocity Technique?}
\author{Ji Wang, and Jian Ge}
\affil{Department of Astronomy, 211, Bryant Space Science Center, University of Florida, Gainesville,
    FL, 32611}
\email{jwang@astro.ufl.edu}

\begin{abstract}
Discovering an Earth-like exoplanet in habitable zone is an important milestone for astronomers in search of extra-terrestrial life. While the radial velocity (RV) technique remains one the most powerful tools in detecting and characterizing exo-planetary systems, we calculate the uncertainties in precision RV measurements considering stellar spectral quality factors, RV calibration sources, stellar noise and telluric contamination in different observational bandpasses and for different spectral types. We predict the optimal observational bandpass for different spectral types using the RV technique under a variety of conditions. We compare the RV signal of an Earth-like planet in the habitable zone (HZ) to the near future state of the art RV precision and attempt to answer the question: How close are we to detecting Earth-like planet in the HZ using the RV technique?

\end{abstract}


\section{Introduction}

Surveys of exoplanets around stars of different spectral types using the RV technique have yielded fruitful discoveries including gas giants, Neptune-like planets and super-Earths. As of 2011 May, there are over 550 detected exoplanets and about 80\% of them were discovered by the radial velocity (RV) technique using ground-based Doppler instruments\footnotemark\footnotetext{http://exoplanet.eu/; http://exoplanets.org/}. However, despite all these discoveries, an Earth-like planet in the HZ has not yet been detected. 

We have not achieved the extremely high RV measurement precisions required in the detection of an Earth-like plaent in the HZ around stars of different masses. At lower end of stellar mass spectrum,~\citet{Endl2006} measured the mean of RV RMS scatter for a sample of 90 M dwarfs, which is 8.3 $\rm{m}\cdot\rm{s}^{-1}$. ~\citet{Bean2010} demonstrated $\sim$5 $\rm{m}\cdot\rm{s}^{-1}$ RV precision for several M dwarfs over a period of 6 months. The intrinsic faintness and lack of an optimal RV calibration source currently limit the improvement of RV precision in the near infrared (NIR). In addition, imperfect stellar template spectra and modeling of telluric lines pose additional problems to extracting RV in the NIR where the flux of an M dwarf peaks~\citep{Bean2010}. For solar type stars, i.e., spectral type of FGK, $\sim$1 $\rm{m}\cdot\rm{s}^{-1}$ RV precision has been routinely achieved~\citep{Mayor2008, Howard2010}. However, stellar activity, stellar granulation and instrumental noise are major obstacles to overcome before detection of Earth-like planets~\citep{Mayor2008}. At the high mass end, a precision of 3$-$6 $\rm{m}\cdot\rm{s}^{-1}$ was demonstrated by ~\citet{Johnson2007} and 6$-$8 $\rm{m}\cdot\rm{s}^{-1}$ RV RMS scatter for stars with planets is found in the search of planets around evolved A type stars~\citep{Bowler2010}. Fast stellar rotation broadens stellar absorption lines and limits the RV precision for main sequence massive stars~\citep{Wright2005}. For evolved massive stars, similar problems exist as those facing solar type stars.

The fundamental photon-noise RV uncertainties have been discussed in several previous papers~\citep{Butler1996,Bouchy2001}. However, only intrinsic properties of stellar spectra are discussed in their works while no detailed calculation of RV uncertainties introduced by the calibration sources. In a recent paper, ~\citet{Reiners2010} considered the uncertainties in the NIR caused by RV calibration sources, i.e., a Th-Ar lamp and an Ammonia gas absorption cell. However, these two RV calibration sources can not be completely representative of the calibration sources used and proposed in current and planned Doppler planet survey in the NIR. For example, there are other emission lamps available in the NIR for RV calibration such as a U-Ne lamp as proposed by~\citet{Mahadevan2010}.  In addition, other gas absorption cells besides the Ammonia cell have been proposed in the NIR~\citep{Mahadevan2009, Valdivielso2010}.  Futhermore, in~\citet{Reiners2010}, the calculation of RV calibration uncertainty of a gas absorption cell assumes a 50 nm band width in $K$ band, and then the uncertainty was applied to other NIR bands, which is purely hypothetical. Therefore, a more comprehensive and detailed study of RV calibration uncertainties is necessary at different observational bandpasses in the NIR in the search of planets around cool stars. On the other hand, in the visible, even though the current RV precision is not limited by the RV calibration source such as a Th-Ar lamp or an Iodine absorption cell, a better understanding of their performances under the photon-limited condition helps us discern a stage in which the RV calibration source becomes the bottle neck as RV precision keeps improving.~\citet{Rodler2011} recently investigated RV precision achievable for M and L dwarfs, but did not quantitatively discussed the influence of RV calibration sources and stellar noise on precision Doppler measurement.

Stellar noise is a significant contributor to RV uncertainty budget, which falls into three categories: p-mode oscillation, spots and plagues, and granulations. P-mode oscillation usually produces an RV signature with a period of several minutes. The oscillation mode has been relatively well studied by previous work (e.g.,~\citet{Carrier2003,Kjeldsen2005}). Exposure time of 10-15 min is proposed in order to smooth the RV signature induced by p-mode oscillation~\citep{Dumusque2011}. Spots and plagues induced RV signal has been discussed by several papers (e.g.,~\citet{Desort2007, Reiners2010, Lagrange2010, Meunier2010}).~\citet{Meunier2010} concluded that the photometric contribution of plages and spots should not prevent detection of Earth-mass planets in the HZ given a very good temporal sampling and signal-to-noise ratio. Granulation is considered to be the major obstacle in detection of Earth planets in the HZ because it produces an RV signal with an amplitude of 8$\sim$10 $\rm{m\cdot s}^{-1}$ based on observation on the Sun~\citep{Meunier2010}. In addition, there is by far no good method of removing the RV noise from this phenomenon.~\citet{Dumusque2011} provided a model of noise contribution in RV measurements based on precision RV observation on stars of different spectral type and at different evolution stages. 

The telluric lines from the Earth's atmosphere are usually masked out in calculations of the photon-limited RV uncertainties in NIR~\citep{Reiners2010, Rodler2011}. Although ~\citet{Wang2011} proposed a method to quantitatively estimate the influence of atmosphere removal residual on precision Doppler measurement using the Dispersed Fixed Delay Interferometer (DFDI) method~\citep{Ge2002,Erskine2003,VanEyken2010}, no attempt has ever been made for precision Doppler measurements using a high-resolution Echelle spectrograph. In practice, the telluric lines are not masked out, but instead modeled and removed. Therefore, a quantitative way of estimating the RV uncertainties produced by the residual of telluric line removal is necessary before we fully understand the performance of an RV instrument. In the visible band, the estimation of telluric line contamination is equally important as higher RV precision is required in the search of lower-mass planets around solar type stars.

We address two basic questions in this paper after considering a variety of factors including stellar spectrum quality, RV calibration precision, stellar noise and atmosphere contamination: 1, which observational bandpass is optimal to conduct precision Doppler measurements for stars of different spectral types; 2, is current RV precision adequate for detecting Earth-like planets in the HZ in the most optimistic scenario, in which the star is the least active and telluric lines are perfectly modeled and removed. The methods and findings of this study will provide insights to the design and optimization of a planned or ongoing precision Doppler planet survey. In addition, it also helps us to access at what stage we are in the search of Earth-like planets in the HZ.

The paper is organized as follows. In \S \ref{sec:method}, we introduce the methodologies of estimating the photon-limited RV uncertainties from difference sources. The findings are presented in \S \ref{sec:result}. Summaries and discussions of this study will be given in \S \ref{sec:Conclusion}.

\section{Methodology}
\label{sec:method}

\subsection{High Resolution Synthetic Spectra}
\label{sec:spec}

Because observed stellar spectra do not have high enough spectral resolution and broad effective temperature coverage, we decide to use high resolution synthetic stellar spectra in the calculation of photon-limited RV uncertainty. For solar type stars, i.e., FGK type stars (3750 K$\leq T_{\rm{eff}}\leq$7000 K), we adopt the spectra with a 0.02 $\AA$ sampling from~\citet{Coelho2005}. For M dwarfs, (2400 K$\leq T_{\rm{eff}}\leq$3500 K),  we use high-resolution (0.005 $\AA$ sampling) synthetic stellar spectra generated by PHOENIX code\citep{Hauschildt1999,Allard2001}. ~\citet{Reiners2010} conducted several comparisons between synthetic spectra generated by PHOENIX and observed spectra in NIR. They concluded that the synthetic spectra are accurate enough for the purpose of simulations. For more massive stars (7000 K$<T_{\rm{eff}}\leq$9600 K), we also use the synthetic spectra with 0.005 $\AA$ sampling generated by PHOENIX. Throughout the paper, we assume a metallicity of solar abundance and a surface gravity $\log g$ of 4.5 for main sequence stars. We assume a Gaussian line spreading function (LSF) which is determined by spectral resolution($\rm{R}$). After an artificial rotational line broadening using a kernel provided by~\citet{Gray1992} and an LSF convolution, we rebin each spectral slice according to 4.0 pixels per resolution element (RE) to generate the a one-dimensional spectrum. In comparison, the sampling rate is 3.2 pixel/RE for HARPS~\citep{Mayor2003} and 3.5 pixel/RE for HIRES~\citep{Vogt1994}.

\begin{figure}
\begin{center}
\includegraphics[width=16cm,height=12cm,angle=0]{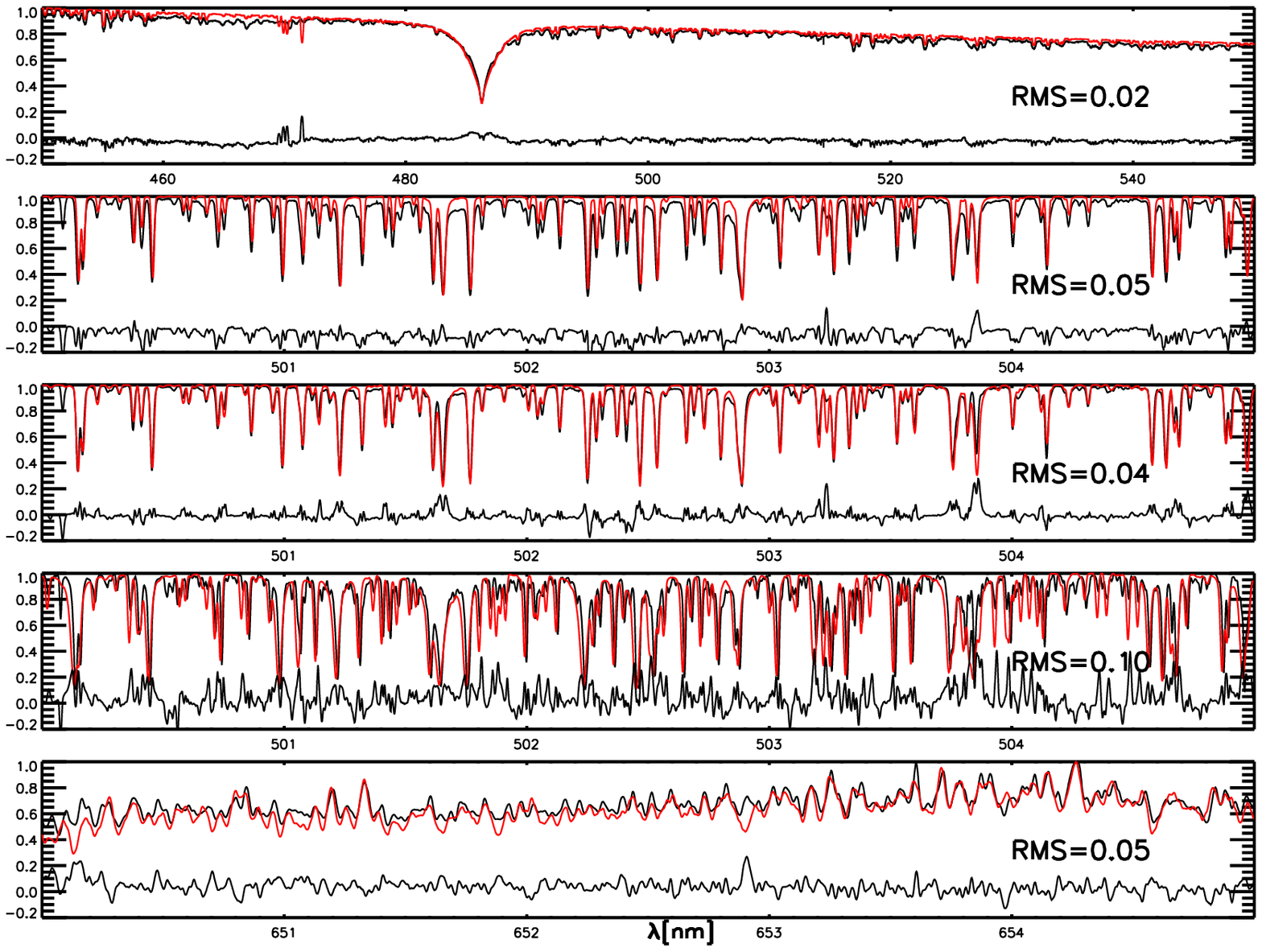}
\caption{Comparisons between synthetic and observed spectra. Black lines represent observed spectra and red lines are synthetic spectra after rotational line broadening and LSF convolution at $\rm{R}$=80,000. The $T_{\rm{eff}}$ and $V\sin i$ are chosen according to the spectral type and line width empirically, they are not necessarily the best-fit parameters for the observed spectra. The chosen $T_{\rm{eff}}$ and $V\sin i$ are, from top to bottom, 9000 K and 80.0 $\rm{km}\cdot\rm{s}^{-1}$ for HD 39060 (A5V), 6250 K and 4.5 $\rm{km}\cdot\rm{s}^{-1}$ for HD 30562 (F8V), 5750 K and 6.0 $\rm{km}\cdot\rm{s}^{-1}$ for HD 14802 (G2V), 4750 K and 4.0 $\rm{km}\cdot\rm{s}^{-1}$ for HD 10361 (K5V), 2900 K and 10.0 $\rm{km}\cdot\rm{s}^{-1}$ for HD 34055 (M6V). The difference between observed and synthetic spectrum is also plotted at the bottom of each panel with RMS of difference.
\label{fig:comp_syn_obs}}
\end{center}
\end{figure}

We compare the synthetic spectra to the observed ones in the visible to ensure that the synthetic spectra are good approximations of observed stellar spectra~\citep{Bagnulo2003}. Comparison in NIR requires carefully removing telluric lines from the observed stellar spectra, which is beyond the scope of this paper. Fig. \ref{fig:comp_syn_obs} shows comparisons of synthetic spectra and the observed high resolution ($\rm{R}$=80,000) stellar spectra from ~\citet{Bagnulo2003}. The comparison spans a wide range of spectral types from M6V to A5V. The synthetic spectrum of an A5V star matches well the an observed one with an RMS of 0.02. As the features in a stellar spectrum increases due to cooler $T_{\rm{eff}}$ and slower stellar rotation, the RMS increases due to an increasing complexity of comparison and imprecise spectral line modeling. The RMS of difference is 0.05 and 0.04 for an F8V and a G2V star. It get worse in the comparison for a K5V star, in which the RMS is 0.10. And the RMS of difference is 0.05 for an M6V star. The results from the comparisons between synthetic and observed spectra indicate the difficulty in modeling the spectra of cool stellar objects. Although not perfect, the synthetic spectra are able to reproduce majority of the features in the observed spectra. Therefore, we decide to use the synthetic spectra in our calculation of RV uncertainty.

\subsection{Spectral Quality Factor $Q$}
\label{sec:Q}

An efficient way to calculate the photon-limited uncertainty in the  Doppler measurements based on a spectral quality factor (Q) was introduced by ~\citet{Bouchy2001}. The Q factor is a measure of spectral profile information within the wavelength region considered for Doppler measurements.  The Q factor is calculated for a series of 10 nm spectral slices in each observational bandpass. According to ~\citet{Bouchy2001}, Q is defined as:
\begin{equation}
\label{eq:q_factor} Q\equiv\Bigg[\frac{{\displaystyle\sum_i
W(i)}}{{N_{e^-}}}\Bigg]^{1/2},
\end{equation}
where $N_{e^-}$ is total number of photons within spectral range, and $W(i)$ is expressed as:
\begin{equation}
W(i)=\frac{\bigg(\frac{\partial
A(i)}{\partial\nu(i)}\bigg)^2\nu(i)^2}{A(i)}.
\end{equation}
In the equations, $i$ is pixel number, $\nu$ is optical frequency, $A$ is a wavelength-calibrated digitalized spectrum. The Q factor is independent of photon flux and represents
extractable Doppler information given the intrinsic stellar spectrum and
instrument spectral resolution. 
The overall RV uncertainty for the entire spectral range is given by~\citep{Bouchy2001}:
\begin{eqnarray}
\label{eq:overall_Doppler} \frac{\delta v_{rms}}{c} & = &
\frac{1}{Q\sqrt{N_{e^-}}},
\end{eqnarray}
according to Equation
(\ref{eq:overall_Doppler}), we can calculate photon-limited RV
uncertainty given the Q factor and the photon count $N_{e^-}$ within a given spectral range.

\subsection{RV Calibration Sources}
\label{sec:RVcal}

RV calibration sources are important in precision Doppler measurements because they not only provide wavelength solutions but also help track drift due to instrument instabilities. The RV uncertainties due to calibrations must be considered if we want to fully understand the performance of an RV instrument. We consider the photon-limited uncertainties introduced by RV calibration sources based  on their spectral quality factors in our paper. Two types of calibration sources have been successfully applied in RV measurements in the visible bands: 1, a Th-Ar emission lamp~\citep{Lovis2007}; 2, an Iodine gas absorption cell~\citep{Butler1996}. Searching for planets in NIR using the RV technique has already been conducted by several groups~\citep{Bean2010,Blake2010,Figueira2010b,Mahadevan2010,Muirhead2011} and several high resolution NIR spectrographs will be put into use in the foreseeable future~\citep{Ge2006, Quirrenbach2010}. We limit the discussions in the emission lamps and gas absorption cells although there are other candidates for RV calibration sources, for examples, laser combs~\citep{Steinmetz2008,Li2008}, which are unfortunately very expensive and not yet readily available, and interferometer calibration sources as proposed by~\citet{Wildi2010} and ~\citet{Wan2010}

In the following discussions, the RV calibration sources are categorized by the observational bandpass in which they are applied. The corresponding wavelength range for each observational bandpass is given in Table \ref{tab:SpecBand}. 

In $B$ band, a Th-Ar lamp is a suitable calibration source. The lines list of a Th-Ar lamp from ~\citet{Lovis2007} is adopted in this paper. Only Thorium lines are used in the calculation because the instability of Argon lines is at the order of $\sim$10 $\rm{m\cdot s}^{-1}$, which is not stable enough for high precision Doppler measurements. A Iodine absorption cell is assumed in $V$ band for RV calibration, a Th-Ar lamp is also considered in this band for comparison. We obtained a high resolution spectrum ($R\geq$200,000) using the Coude Spectrograph at Kitt peak for a Iodine cell with a 6-inch light path at 60 $^\circ\rm{C}$. Note that an iodine cell spectrum is superimposed on a stellar spectrum~\citep{Butler1996}, the S/N of RV calibration is thus determined by the S/N of the continuum of a stellar spectrum. This case is called Superimposing in this paper. Howerver, for very stable instruments, there are other ways of calibrating the non-stellar drift including spatial~\citep{Mayor2003} and temporal approaches~\citep{Lee2011}. In a spatial approach, the light from a star and a Th-Ar lamp is fed onto nearby but different parts of CCD by two separate fibers (We call this case Non-Common Path in the paper). In Bracketing method, on the other hand, RV calibrations are conducted right before and after a stellar exposure in a temporal approach. In both cases, the S/N of RV calibration is not dependent on stellar flux. The disadvantage is, however, the light  from a calibration source does not pass through the instrument in exactly the same path or at the same time as the light from a star. We choose a Th-Ar lamp as the RV calibration source in $R$ band, where strong Argon lines exist that saturate the CCD. Since we exclude Argon lines in RV calibration uncertainty calculation, a more practical result when Argon lines are considered is expected to be worse unless a CCD with higher dynamic range is used. 

In $Y$ and $J$ band, a U-Ne emission lamp is proposed by ~\citet{Mahadevan2010}, we use a lines list of Uranium provided by Stephen Redman~\citep{Redman2011}. In $H$ band, a series of absorption cells is proposed by ~\citet{Mahadevan2009}, in which a mixture of gas cells including $H^{13}C^{14}N$, $^{12}C_2H_2$, $^{12}CO$, and $^{13}CO$ creates a series of absorption lines that spans over 120 nm of the $H$ band. ~\citet{Bean2010} demonstrated that an Ammonia absorption cell is a good candidate for calibration source in $K$ band. Therefore, we assume an Ammonia cell in the calculation of RV calibration uncertainty in the $K$ band. ~\citet{Valdivielso2010} proposed a gas absorption cell with the mixture of acetylene, nitrous oxide, ammonia, chloromethanes, and hydrocarbons covering most of the $H$ and $K$ bands. We do not consider this cell in our paper since a detailed lines list of the cell is not available. 

\subsection{Stellar Noise}
\label{sec:StellarNoise}

Stellar noise is a significant contributor to RV uncertainty budget, therefore we devote the following part to discuss a method of quantifying its influence on precision Doppler measurement. Granulation is considered to be the major obstacle in detection of Earth planets in the HZ because it produces an RV signal with an amplitude of 8$\sim$10 $\rm{m\cdot s}^{-1}$ based on observation on the Sun~\citep{Meunier2010}. In addition, there is by far no good method of removing the RV noise from this phenomenon.~\citet{Dumusque2011} provided a model of noise contribution in RV measurements based on precision RV observation on stars of different spectral type and at different evolution stages. We adopt this model and quantify the RV uncertainty contribution of granulation based their measurement of three stars, i.e., $\alpha$ Cen A (G2V), $\tau$ Ceti (G8V), and $\alpha$ Cen B (K1V). The sum of three exponentially decaying functions represents a power spectrum density function with contributions from granulation, meso-granulation and super-granulation, using the values given in Table 2 from~\citet{Dumusque2011}. An RV RMS error due to granulation is then calculation based on Equation (6) in their paper assuming a 100-day (300-day) consecutive observation for K (G) type star with an optimal strategy found in the paper, i.e., three measurements per night of 10 min exposure each, 2 h apart. The total length of consecutive observation is roughly in accordance with the orbital period of a planet in the HZ. We find that the RV RMS error due to granulation is 0.55, 1.05 and 1.05 $\rm{m\cdot s}^{-1}$ for a K1V, G8V and G2V star respectively. These number are going to be used later in this study to estimate a total RV uncertainty. 

Detailed study of RV uncertainty induced by stellar noise has so far been limited in K and G type stars due to practical concerns such as stellar photon flux and stellar activity. Despite their intrinsic faintness and relative higher level of stellar activity due to fast rotation and deep convection zone, M dwarfs are among primary targets in search of planets in the HZ. The RMS fitting error of orbit of GJ 674 b~\citep{Bonfils2007} is 0.82 $\rm{m\cdot s}^{-1}$ after RV noise due to a stellar spot is modeled and removed, providing an good target for Earth-like planet search with an upper limit of other stellar noise contribution of 0.82 $\rm{m\cdot s}^{-1}$, if we interpret the RMS fitting error is due to an combination of instrument instability, photon-noise and other source of stellar noise. We adopt the model proposed by~\citet{Dumusque2011} to estimate the RV RMS error due granulation phenomenon for an M dwarf using the parameters for a K or G star. Aware of the caveat  of different stellar type, we find the RMS error is 0.52, 1.07 and 1.04 $\rm{m\cdot s}^{-1}$ using the parameters for a K1V, G8V or G2V star. 50-day consecutive observation is assumed with an optimal strategy described in~\citet{Dumusque2011}. The theoretical calculation of granulation-induced RV RMS error is worse than observation of GJ 674 b using parameters for G stars, suggesting the parameters for G stars are not representative of optimistic scenario in observation of M dwarfs. Therefore, we use 0.52 $\rm{m\cdot s}^{-1}$, which is a result of using the parameters for K stars, as an estimation of granulation-induced RV RMS error for an M dwarf in an optimistic case. 

\subsection{Telluric Lines Contamination}
\label{sec:Telluci}

Ground-based observations are prone to contamination by telluric
lines. Precision Doppler measurements in the NIR requires a significant level of disentanglement of stellar absorption
lines and telluric lines. As RV precision keeps improving, Doppler measurements in the visible band such as $B$, $V$ and $R$ band should also consider telluric lines, because the contamination of them will no longer be negligible. The quantification of telluric line contamination has been discussed by~\citet{Wang2011} in the context of Dispersed Fixed Delay Interferometer method~\citep{Ge2002,Erskine2003,VanEyken2010}, here we present a generalization of the method for the case of Echelle spectrograph, which is a more conventional application. Atmospheric transmission (AT) is calculated by a service provided by $spectralcalc.com$ based on a method described in~\citet{Gordley1994}. 
 
The following equation describes the flux distribution on a CCD detector if
telluric absorption lines are considered:
\begin{eqnarray}
F(\nu,y) & = & \bigg[\frac{S_0(\nu)}{h\nu}\times AT(\nu)\bigg]\otimes LSF\nonumber \\
   & = & \bigg[\frac{S_0(\nu)}{h\nu}\times (1-\alpha\times AA(\nu))\bigg]\otimes LSF\nonumber \\
   & = & \bigg[\frac{S_0(\nu)}{h\nu}\bigg]\otimes LSF+\bigg[-\frac{S_0(\nu)}{h\nu}\times \alpha\times AA(\nu)\bigg]\otimes LSF \nonumber \\
   & = & F_S(\nu,y)+F_N(\nu,y),
\label{eq:B_atm}
\end{eqnarray}
where $S_0(\nu)$ is stellar energy flux, which is converted into photon flux by being divided by $h\nu$, $AT$ is the atmospheric transmission function, $AA$ is the
atmospheric absorption function, and $\alpha$ is a parameter describing the level of telluric line removal as a first-order estimation.

In Equation (\ref{eq:B_atm}),
photon flux distribution on the detector, $F$, is comprised of a
signal component $F_S$ and a noise component $F_N$. Ideally, we
require that the detector flux change, $\delta F$, is entirely due to the
stellar RV change $\delta v_S$. However, $\delta F$ is also partly
induced by telluric line shift $\delta v_N$ resulting from random
atmospheric motions. Therefore, both $\delta v_S$ and $\delta v_N$
contribute to $\delta F$. We have two sets of RV measurements,
$\delta v_S+\sigma(0,\delta v_{rms,S})$ for stellar RV and $\delta
v_N+\sigma(0,\delta v_{rms,N})$ for RV induced by the Earth's atmosphere, where
$\sigma(0,\delta)$ represents random numbers following a gaussian
distribution with a mean of 0 and a standard deviation of $\delta$. $\delta v_{rms,S}$ is the photon-limited measurement error for component $F_S$ and $\delta v_{rms,N}$ is the photon-limited measurement error for component $F_N$. We weigh the final RV measurement with the inverse square of photon-limited
RV uncertainties of these two components, which is expressed by the
following equation:
    \begin{equation}
    \label{eq:simple_example_tulleric}
\delta v=\frac{(\delta v_S+\sigma(0,\delta
v_{rms,S}))\cdot{\delta v_{rms,S}^{-2}}+(\delta
v_N+\sigma(0,\delta v_{rms,N}))\cdot{\delta
v_{rms,N}^{-2}}}{{\delta v_{rms,S}^{-2}}+{\delta
v_{rms,N}^{-2}}},
\end{equation}
In practical Doppler measurements, $\delta v_S$ consists two components, stellar RV and Earth's barycentric RV. Depending on the position of the Earth in its orbit, there is an offset between $\delta v_S$ and $\delta v_N$, which is the Earth's barycentric velocity. The Earth's barycentric motion has a semi-amplitude of 30 $\rm{km}\cdot\rm{s}^{-1}$. Statistically, observed star has an annually-varying RV with a semi-amplitude of on-average 21.21 $\rm{km}\cdot\rm{s}^{-1}$. We artificially shift a stellar spectrum by an amount less than 21.21 $\rm{km}\cdot\rm{s}^{-1}$ in order to generate a offset between stellar spectrum and $AA$ spectrum. $\delta v_{rms,S}$ and $\delta v_{rms,N}$ are then calculated for $F_S$ and $F_N$. We choose the median of $\delta v_{rms,N}$ to represent a typical $\delta v_{rms,N}$ value from calculations based on different input barycentric velocities. We further assume that observed star has a constant RV (i.e., no differential RV), and $F_N$ has an RV fluctuation with an RMS of $\delta v_{N,ATM}$ because of the Earth's  turbulent atmosphere. The measured RV
uncertainty $\delta v$ is equal to:
    \begin{equation}
    \label{eq:simple_example_tulleric_2}
\delta v_{rms}=\frac{(\delta v_{rms,S})\cdot{\delta
v_{rms,S}^{-2}}+{(\delta v_{N,ATM}^2+\delta
v_{rms,N}^2)^{1/2}}\cdot{\delta v_{rms,N}^{-2}}}{{\delta
v_{rms,S}^{-2}}+{\delta v_{rms,N}^{-2}}},
\end{equation}
In reality, RV uncertainty of $F_N$ is not dominated by photon-noise, instead, it is dominated by atmospheric behaviors such as wind, molecular column density change, etc.~\citet{Figueira2010} used HARPS archive data and found
that $O_2$ lines are stable to a 10 $\rm{m\cdot s}^{-1}$ level over 6 years. However,
long term stability of telluric lines (over years) becomes worse if we take into
consideration other gas molecules such as $H_2O$ and $CO_2$. The uncertainty induced by atmospheric telluric lines is transferred to $\delta v_{rms}$ via Equation (\ref{eq:simple_example_tulleric_2}). In order to calculate the final RV uncertainty, $\delta v_{rms}$, we need to calculate photon-limited RV uncertainty $\delta v_{S,rms}$ and $\delta v_{N,rms}$ according to Equation (\ref{eq:overall_Doppler}), in which two terms need to be calculated: Q and $N_{e^-}$. The spectral quality factors ($Q_S$ and $Q_N$) for the two components ($F_S$ and $F_N$) from equation (\ref{eq:B_atm}) are calculated based on Equation (\ref{eq:q_factor}).  $N_{e^-,S}$ and $N_{e^-,N}$, the photon flux of $F_S$ and $F_N$ are calculated based on stellar type, magnitude, exposure time, instrument specifications and telluric absorption properties. Note the ratio of $N_{e^-,S}$ and $N_{e^-,N}$ remains constant as long as atmospheric absorption stays unchanged because telluric line absorption is imprinted on the stellar spectrum. 

%Three examples are given here to represent different levels of 
%telluric line contamination in the precision RV measurements: 1, if $\delta v_{rms,S}\gg\delta v_{rms,N}$, then $\delta v_{rms}={(\delta v_{N,ATM}^2+\delta
%v_{rms,N}^2)^{1/2}}$, and the RV measurement is dominated by random atmospheric motions, this approximation applies in a wavelength region with dense telluric line distribution;
%2, if $\delta v_{rms,S}\ll\delta v_{rms,N}$ (i.e., in very transparent atmosphere windows where few telluric lines exists), then $\delta
%v_{rms}=\delta v_{rms,S}$, and the RV uncertainty is limited by stellar
%photon noise; 3, for an intermediate situation, if $\delta v_{rms,S}$ and $\delta
%v_{rms,N}$ are identical, then we apply the same weight on both RV
%measurements, $\delta v_{rms}={[\delta v_{rms,S}+{(\delta
%v_{N,ATM}^2+\delta v_{rms,N}^2)^{1/2}}]/2}$.

The method described above provides a quantitative way of answering the questions such as: 1, how the RV uncertainty is correlated with different levels of residual of telluric line removal; 2, what  the contribution of RV uncertainty due to telluric contamination is in the final RV error budget in each different observational bandpass. 
%As a first order estimation, we consider the simplest case in which the centroids of telluric line remain unshifted and the only uncertainty is the depth of a telluric line. 
%In the process of telluric
%lines modeling, we remove
%telluric lines from the observed spectrum of a science target. The depths of
%the telluric lines are consequently reduced by a certain fraction
%depending upon the level of removal, which effectively
%reduces $AA$ (Atmospheric Absorption). Therefore, Equation (\ref{eq:B_atm}) is rewritten as follows:
%\begin{equation}
%F(\nu,y)=F_S(\nu,y)+\alpha\times F_N(\nu,y),
%\label{eq:B_atm_remov}
%\end{equation}
%where $\alpha$ is a parameter describing the level of telluric line removal. If no telluric line removal is performed, then $\alpha$ is equal to 1; if telluric lines are completely removed from observed stellar spectrum, then $\alpha$ is equal to zero. A reduced noise component, $\alpha\times F_N$, means less photon contribution and thus less uncertainty induced by atmospheric behaviors. 

\section{Results}
\label{sec:result}

\subsection{RV Calibration Uncertainty}
\label{sec:RVcalUn}

RV calibration sources are used to track the drift that is not caused by the stellar reflex motion due to an unseen companion. A emission lamp or a gas absorption cell is usually used for such purpose. We calculate the RV uncertainties brought by the calibration sources themselves based on their spectral properties. The RV calibration sources in different observational bandpasses are discussed in \S \ref{sec:RVcal}. For gas absorption cells, we assume a continuum level of 30,000 ADU (within the typical linear range) on a CCD with a 16-bit dynamic range, which corresponds to a S/N of 425 if the gain is at 6 electron/ADU. For a emission lamp, we assume that the strongest line in the spectral region has a peak flux of 30,000 ADU. Note that 4 pixels per resolution element is assumed throughout the paper. Fig. \ref{fig:Un_Cal_Res} shows the RV calibration uncertainties as a function of observational bandpass at different spectral resolutions. Note that there are currently two successful calibration sources in $V$ band, i.e., a Th-Ar lamp (asterisk) and an Iodine cell (square). Therefore, both are considered and plotted for $V$ band in Fig. \ref{fig:Un_Cal_Res}. The results shown in Fig. \ref{fig:Un_Cal_Res} are also summarized in Table \ref{tab:RV_cal}. Gas absorption cells usually offer higher calibration precision than emission lamps because of denser lines distribution and on-average higher S/N. However, this conclusion depends on the calibration methods, it is usually the case for Non-Common Path and Bracketing method while it is not always true in Superimposing scheme, which will be discussed later in this section.

\subsection{Optimal Spectral Band For RV Measurements}
\label{sec:OptiBand}

The optimal observational bandpass for precision RV measurements depends on the quality of a stellar spectrum (Q factor), photon flux (S/N), RV calibration uncertainty, the severity of telluric line contamination and other factors. We will consider different situations in the following discussion. We assume a S/N (per pixel) of 100 at the center of $Y$ band (i.e., $\lambda$=1020 nm) at $\rm{R}$=60,000, the S/N in other observational bandpass varies with stellar spectral energy distribution (SED) and spectral resolution accordingly. The S/N reported in this paper is at the center of each observational bandpass (see Table \ref{tab:SpecBand}) unless otherwise specified. We will investigate the optimal observational bandpass for precision Doppler measurements given the same exposure time, the same telescope aperture and the same instrument throughput (independent of wavelength). 

\subsubsection{Stellar Spectral Quality}
\label{sec:OptiBand_Star}

\begin{figure}
\begin{center}
\includegraphics[width=16cm,height=12cm,angle=0]{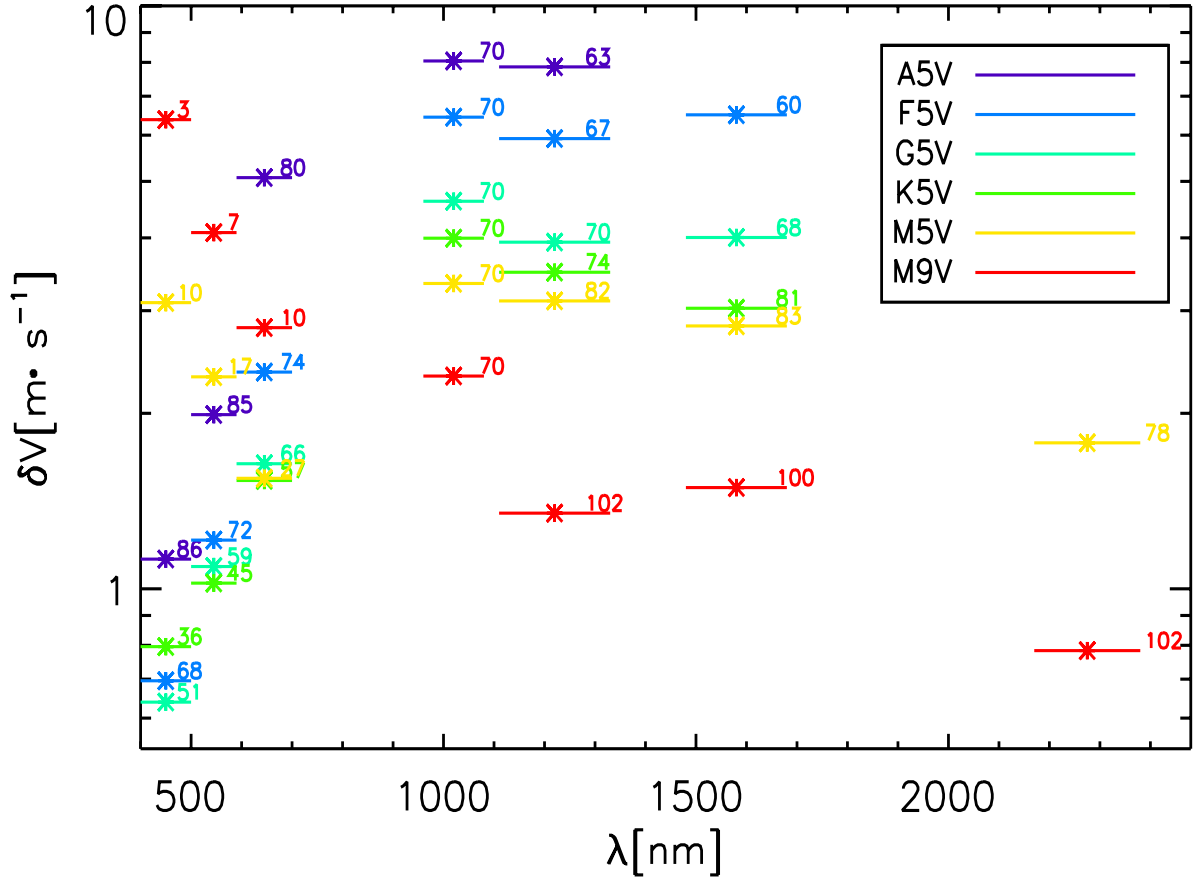}
\caption{RV precision ($R$=120,000) based on spectral quality factor as a function of observational bandpass for different spectral types from M9V to A5V (color-coded). Average S/N per pixel is also shown in the plot, see Table \ref{tab:RV_spec} for results at other spectral resolutions. K band RV uncertainties are not calculated for stars with $T_{\rm{eff}}$ higher than 3500 K because they are usually observed in the visible band at current stage.
\label{fig:RV_wav_120000}}
\end{center}
\end{figure}

We start with the simplest case in which the RV uncertainty is only determined by the stellar spectral quality factor and the SED of a star. In other words, the RV calibration source is perfect and no uncertainty is introduced when calibrating out the non-stellar drift. In addition, telluric lines are perfectly removed from the observed stellar spectrum. Table \ref{tab:RV_spec} summarizes the obtainable RV precisions and the S/Ns at three different spectral resolutions, i.e., 20,000, 60,000 and 120,000. An example of $\rm{R}=120,000$ is plotted in Fig. \ref{fig:RV_wav_120000}. We find the optimal observational bandpass is $B$ band for a wide range of spectral types from K to A. The optimal observational bandpass for an M dwarf is either in $R$ band or in $K$ band. More specifically, $R$ band is optimal for an early-to-mid-type M dwarf while $K$ band for an late-type M dwarf. The finding remains the same for a wide range of spectral resolutions from 20,000 to 120,000. The RV uncertainty for another spectral type or at a different S/N can be obtained by either interpolation or scaling based on the results in Table \ref{tab:RV_spec}. 

\subsubsection{Stellar Spectral Quality + Stellar Rotation}
\label{sec:OptiBand_Star_rot}

\begin{figure}
\begin{center}
\includegraphics[width=16cm,height=12cm,angle=0]{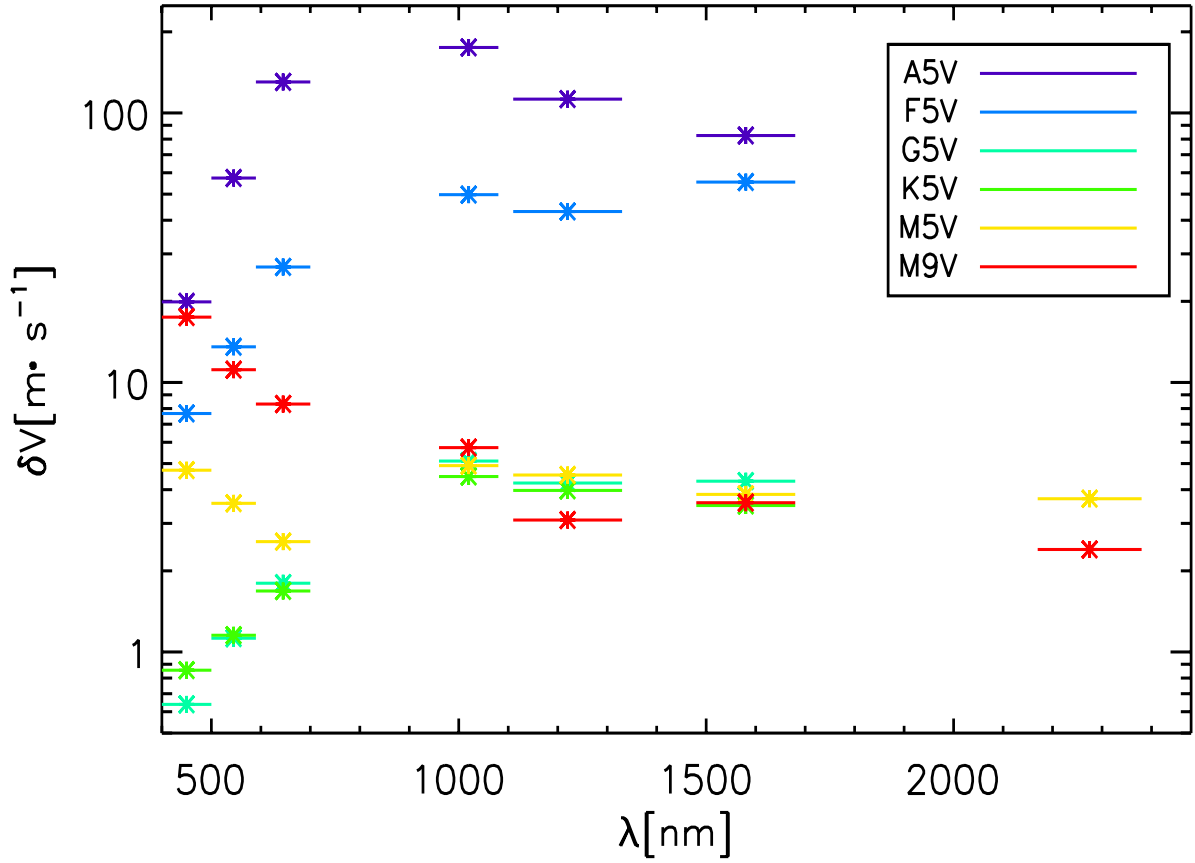}
\caption{RV precision ($R$=120,000) based on spectral quality factor and typical stellar rotation as a function of observational bandpass for different spectral types from M9V to A5V (color-coded). Average S/N per pixel is the same as shown in Fig. \ref{fig:RV_wav_120000}.
\label{fig:RV_wav_rot_120000}}
\end{center}
\end{figure}

Stellar rotation broadens the absorption lines in a stellar spectrum, resulting in less Doppler sensitivity. It is therefore necessary to consider the stellar rotation in the discussion of photon-limited RV uncertainty. Typical values of rotation velocities of different spectral types are obtained based on the measurement results from ~\citet{Jenkins2009} for M dwarfs and ~\citet{Valenti2005} for FGK stars. In addition, typical rotational velocities for early type stars such as A stars are extrapolated from values of solar type stars. Table \ref{tab:TypeTeff} summarizes the spectral types and the corresponding $T_{\rm{eff}}$ and V$\sin i$ used in the paper. A trend of increasing  V$\sin i$ is seen as spectral type moves either to early type end (F and A) or late type end (M). After considering typical stellar rotation velocities for different spectral types (as shown in Fig. \ref{fig:RV_wav_rot_120000}, $\rm{R}$=120,000), F and A stars are not suitable targets for precision Doppler measurements because of their intrinsic high stellar rotation. M dwarfs RV uncertainties are getting worse than non-rotating case, but 2$-$5 $\rm{m}\cdot\rm{s}^{-1}$ RV precision are expected in optimal cases, i.e., $R$ band for M5V and $K$ band for M9V. For K and G stars, sub $\rm{m}\cdot\rm{s}^{-1}$ precision is reached under photon-limited condition even after considering typical stellar rotation.

\subsubsection{Stellar Spectral Quality + RV Calibration Source}
\label{sec:OptiBand_Star_RVcal}

\begin{figure}
\begin{center}
\includegraphics[width=16cm,height=12cm,angle=0]{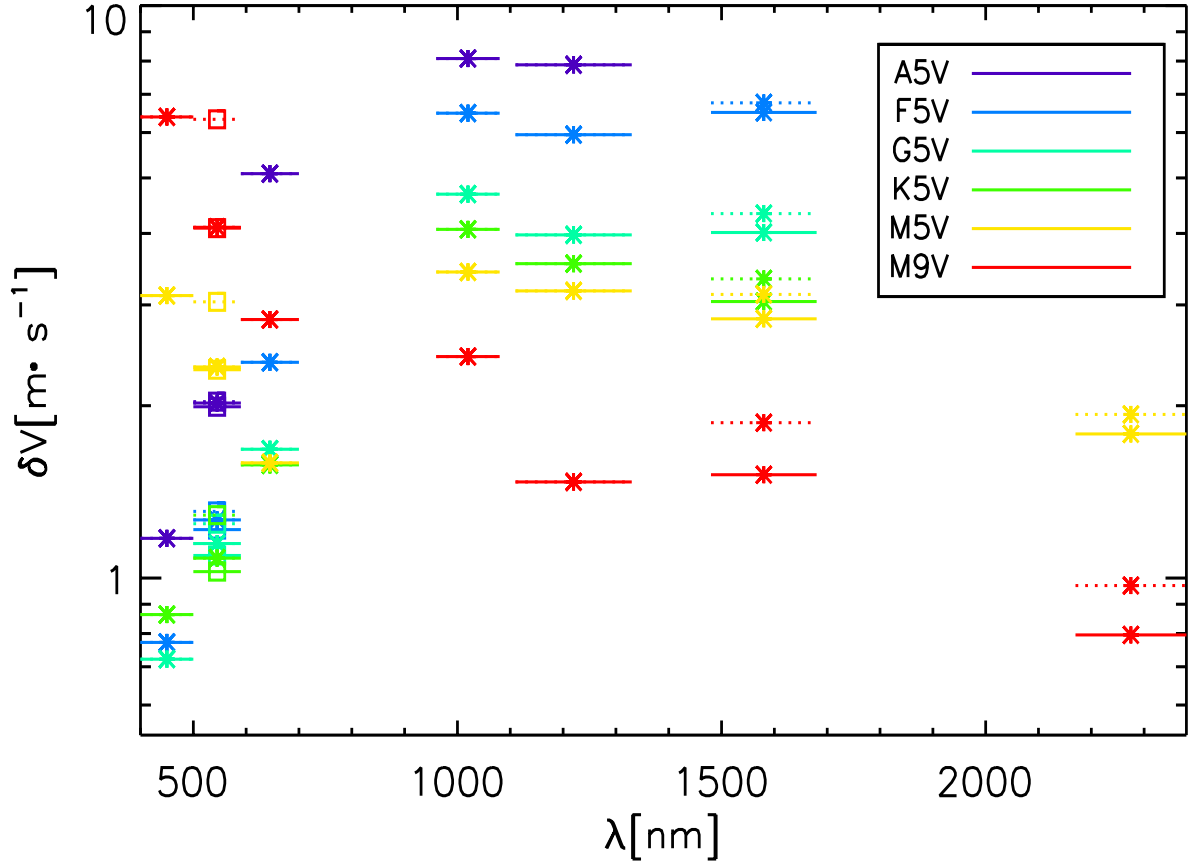}
\caption{RV precision ($R$=120,000) based on spectral quality factor and RV calibration uncertainties as a function of observational bandpass for different spectral types from M9V to A5V (color-coded). Average S/N per pixel is the same as shown in Fig. \ref{fig:RV_wav_120000}. Squares in $V$ band represent Iodine cell method and asterisks in $V$ band represent Th-Ar lamp method. Dotted lines show result from Superimposing cases and solid lines for Non-Common Path and Bracketing cases. Refer to Table \ref{tab:RV_cal} for RV calibration sources in different observational bandpasses.
\label{fig:RV_wav_cal_120000}}
\end{center}
\end{figure}

The above RV precisions considering typical stellar rotation broadening are good indicators for RV planet surveys in which a population of stars is observed with a distribution of stellar rotations. However, in the search for an Earth-like planet, a different approach is taken in which stars with favorable properties for Doppler measurements are assigned higher priority in observation. The properties usually include slow stellar rotation and low stellar activity. Therefore, we will reduce stellar rotation in the following discussion since we emphasize discovery of an Earth-like planet. After considering the uncertainties brought by an RV calibration source, RV uncertainties in Fig. \ref{fig:RV_wav_120000} degrade to those in Fig. \ref{fig:RV_wav_cal_120000}  ($\rm{R}=120,000$). Two scenarios of calibration are considered: Superimposing (Dotted), and Non-Common Path and Bracketing (Solid). The difference between these two is whether the S/N depends on the stellar flux. In the Superimposing case, because the absorption cell is in the light path of the stellar flux, the continuum of the resulting Iodine absorption spectrum is determined by the the continuum flux of a star. Consequently, the RV calibration uncertainty is strongly dependent on the incoming stellar flux. In the comparison of the two cases in Fig. \ref{fig:RV_wav_cal_120000}, we see the Non-Common Path and the Bracketing methods always introduce less uncertainty in RV calibration than the Superimposing method. The major reason for that is the S/N in the former case may be optimized by adjusting the source intensity (Non-Common Path) or the exposure time (Bracketing). The main conclusion about optimal observational bandpass from \S \ref{sec:OptiBand_Star} remains unchanged.

\subsubsection{Stellar Spectral Quality + RV Calibration Source + Atmosphere}
\label{sec:OptiBand_Star_RVcal_Atmosphere}

The optimal band for Doppler measurements is in the NIR (K band in particular) for stars with spectral types later than M5 from previous discussions in this paper. However, one important element is missing in the discussion, which is the contamination from the telluric lines in the Earth's atmosphere, which is a severe problem in the NIR observation. The quantitative analysis of telluric line contamination is introduced in \S \ref{sec:Telluci} and we apply that method in estimating the RV uncertainty brought by the telluric contamination. We confine our discussions for late-type M dwarfs since NIR observation does not gain advantage for other spectral types earlier than M5V. 

\begin{figure}
\begin{center}
\includegraphics[width=16cm,height=12cm,angle=0]{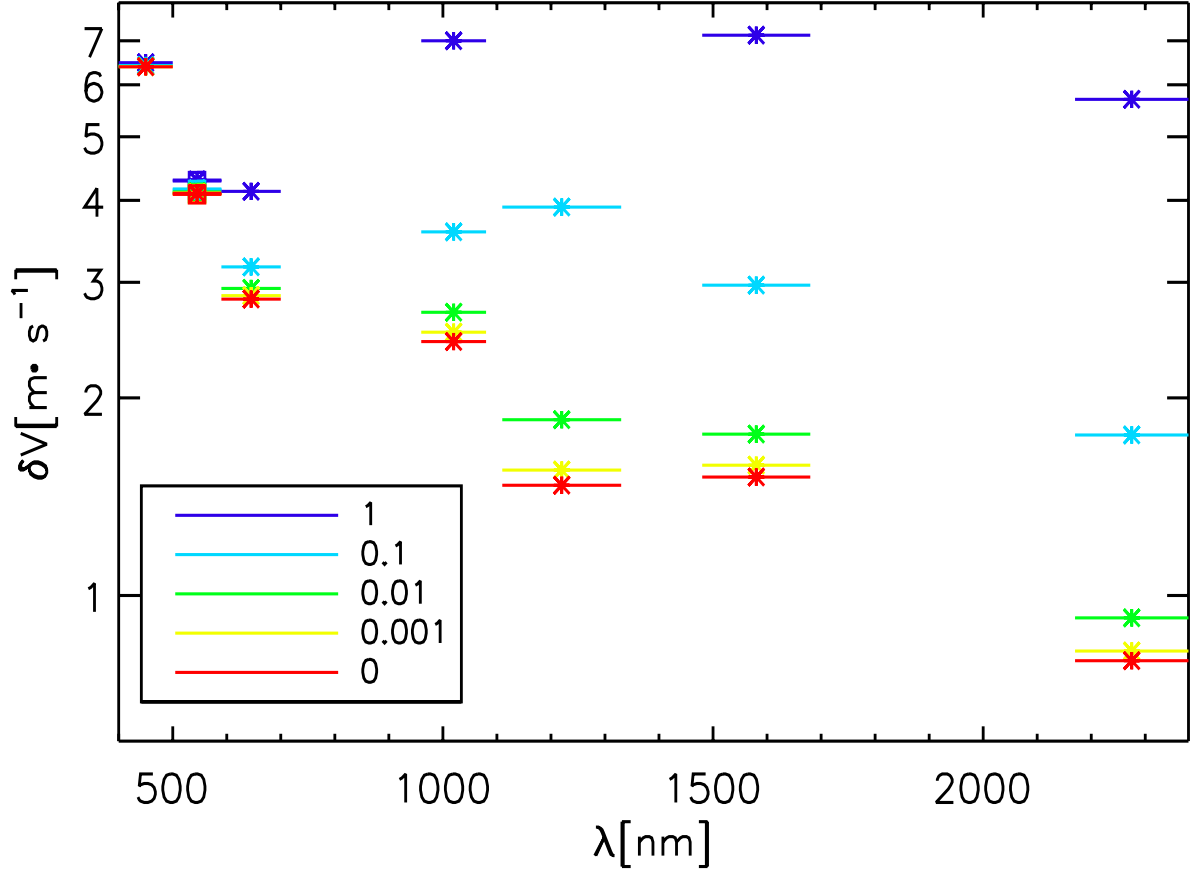}
\caption{RV precision ($R$=120,000) considering spectral quality factor, RV calibration uncertainties and telluric contamination for an M9V star as a function of $\alpha$, i.e., telluric line removal level (color-coded). 1 indicates no telluric line removal and 0 indicates complete removal of telluric lines.  Bracketing RV calibration is assumed for the results shown in the plot. Squares in $V$ band represent Iodine cell method and asterisks in $V$ band represent Th-Ar lamp method. Refer to Table \ref{tab:RV_cal} for RV calibration sources in different observational bandpasses.
\label{fig:RV_wav_cal_atm_120000}}
\end{center}
\end{figure}

Fig. \ref{fig:RV_wav_cal_atm_120000} shows an example of how RV uncertainty for an M9V star changes with observational bandpass under different values of $\alpha$ (i.e., level of telluric line removal, see Equation (\ref{eq:B_atm})). 1 indicates no telluric line removal and 0 indicates complete removal of telluric lines (see Equation (\ref{eq:B_atm})). RV fluctuation of 10 $\rm{m}\cdot\rm{s}^{-1}$ due to random atmospheric movement is assumed in the calculation. Bracketing RV calibration is assumed in the calculation. There are several points worth noting in this plot: 1, different observational bandpasses are affected differently by telluric lines, the significance of telluric line contamination is indicated by the span of RV uncertainties at different $\alpha$ values. For example, $B$ band is the least affected by telluric lines because the RV uncertainties in $B$ band at different levels of telluric line removal remain roughly the same, while $J$, $H$ and $K$ bands suffer severe telluric line contamination because any small change of $\alpha$ results in significant change of RV uncertainty. 2, If there is no attempt of removing telluric lines from observed stellar spectrum (purple in Fig. \ref{fig:RV_wav_cal_atm_120000}), there is no advantage in observing late-type M dwarfs in NIR, RV uncertainty is dominated by Earth's atmosphere behavior in the NIR. In this case, the optimal band is $V$ and $R$ band. Only when $\alpha\leq$0.01, i.e., more than 99\% telluric line strength is removed, the advantage of observing late-type M dwarfs in the NIR becomes obvious, at a factor of $\sim$3 improvement. 

In practice, there have been several examples in which telluric line modeling and removal is demonstrated to be successful.~\citet{Vacca2003} achieved maximum deviations of less than 1.5\% and RMS deviations of less than 0.75\% with $R$=2000 and S/N$\geq$100 using a telluric standard star nearby the science target star.~\citet{Bean2010} has shown that the RMS deviation is as low as 0.7\% after using a 3-component model (Stellar spectrum, telluric absorption and Ammonia absorption) to fit an observed NIR spectrum. In both cases, an $\alpha$ value of better than 0.01 has been demonstrated showing great potential of precision Doppler measurement in the NIR band. 

\begin{figure}
\begin{center}
\includegraphics[width=16cm,height=12cm,angle=0]{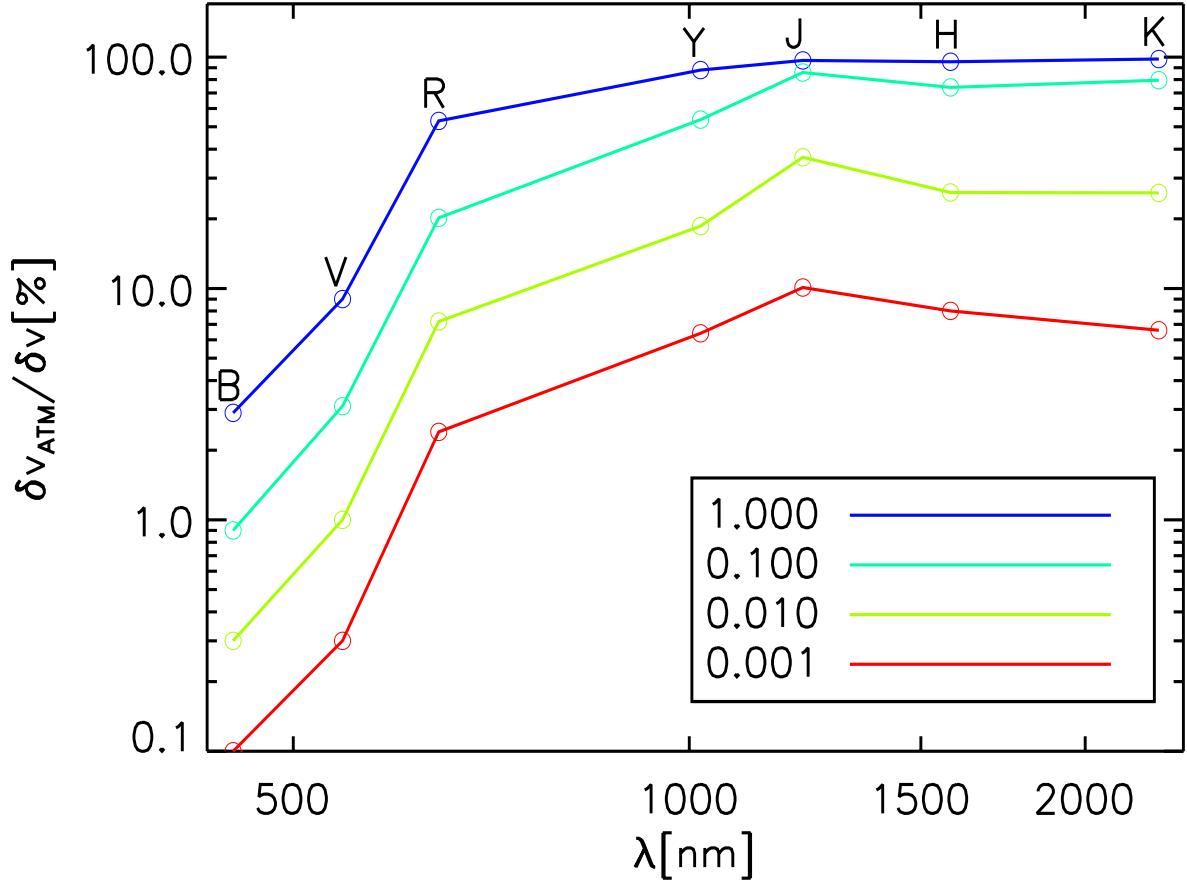}
\caption{The percentage contribution of RV uncertainty induced by telluric contamination as a function of observational bandpass. Different $\alpha$ values are indicated by colors. 1 indicates no telluric line removal and 0.001 indicates 99.9\% effective removal of telluric lines.
\label{fig:Frac_Wav}}
\end{center}
\end{figure}

Fig. \ref{fig:Frac_Wav} shows the percentage contribution of RV uncertainty introduced by telluric contamination at different $\alpha$ values. If no telluric line removal is performed, the RV uncertainty in the NIR is dominated by those caused by telluric contamination, i.e., the percentage contributions are more than 87.9\% in $Y$, $J$, $H$ and $K$ band. In comparison, the percentage contribution of telluric contamination induced RV uncertainty is 2.9\%, 3.1\% and 53.0\% in $B$, $Y$ and $R$ band respectively. As $\alpha$ decreases, i.e., more strength of telluric lines is removed, less RV uncertainty is contributed to the final RV uncertainty budget. However, there is still a significant fraction (more than 70\%) of RV uncertainty contributed by telluric contamination in $J$, $H$ and $K$ band even after 90\% of telluric line strength is removed. The percentage contribution drops below 10\% throughout considered observational bandpasses when more than 99.9\% strength is removed. To sum up the discussion, RV uncertainty is dominated by telluric contamination in the NIR band. Therefore, telluric line removal in the NIR is a necessary step to reduce the telluric contamination and extract more of Doppler information intrinsically carried by a stellar spectrum.

\subsubsection{Comparisons to Previous Work}
\label{sec:Comp_Prev}

There are works that have been previously done in attempts to understand the fundamental photon-limited RV uncertainties based on high resolution synthetic stellar spectra. ~\citet{Bouchy2001} calculated Q factors for a set of synthetic stellar spectra for solar type dwarf stars. We restrict the comparison to spectra with the same turbulence velocity ($V_t$). Since the spectra for solar type stars in our study have a $V_t$ of 1.0 $\rm{km}\cdot\rm{s}^{-1}$, we only compare the results from spectra with $V_t$ of 1.0 $\rm{km}\cdot\rm{s}^{-1}$ in ~\citet{Bouchy2001}. Table \ref{tab:Comp_Bouchy} summarizes a comparison of our results to those from ~\citet{Bouchy2001}. The Q factors from our study are generally 10$-$15\% lower if no stellar rotation is considered, i.e., $V\sin i$=0 $\rm{km}\cdot\rm{s}^{-1}$. It may be due a different sampling rate in the synthetic spectra, 0.005 $\AA$ in ~\citet{Bouchy2001} and 0.02 $\AA$ in our paper. More fine features are seen in a spectrum with higher sampling rate and thus more Doppler information is contained. At low stellar rotation rate ($V\sin i$=4 and 8 $\rm{km}\cdot\rm{s}^{-1}$), our results agree with theirs within 6\%, which is improved compared to non-rotating case because the fine features are smoothed out by stellar rotation. For fast rotators, i.e., $V\sin i$=12 $\rm{km}\cdot\rm{s}^{-1}$, 10\% difference is seen in the worst case, for which a different limb-darkening value might be responsible. 

~\citet{Reiners2010} investigated the precision that can be reached in RV measurements for stellar objects cooler than solar type stars in the NIR. The treatment of telluric lines in their calculations was to block the regions where the telluric absorption is over 2\% and 30 $\rm{km}\cdot\rm{s}^{-1}$ in the vicinity. Following the method described in their paper, we calculated the fraction of the wavelength range affected by telluric contamination in $V$, $Y$, $J$ and $H$ band, the results are 2.4\%, 22.7\%, 60.0\% and 50.6\%. In comparison, the results are 2\%, 19\%, 55\% and 46\% for $V$, $Y$, $J$ and $H$ band in their paper. The difference may be caused by a different atmospheric absorption used in calculation.

Both~\citet{Reiners2010} and we reach the same conclusion that NIR RV measurements start to gain advantage over visible bands for mid-to-late-type M dwarfs. However, we predict that $Y$ and $H$ band are similar in terms of  giving the highest RV precision among $V$, $Y$, $J$ and $H$ bands, while it is found in their paper that $Y$ band is the optimal band considering stellar spectrum quality and telluric line masking. Note that we adopt the definition of $V$, $Y$, $J$ and $H$ bands according to~\citet{Reiners2010} in comparisons. Table \ref{tab:Comp_Reiners} summarizes our calculations of RV precision can be reached for an M9V star ($T_{\rm{eff}}=2400 K$) in comparison to the results in their paper as well as the S/N obtained in each observational bandpass. In further examination, we compare our results in $J$ and $H$ band and find that RV precisions in $H$ band are in general better than those in $J$ band. It is explained by the wider wavelength coverage and richer absorption features in $H$ band for an M9V star. On the contrast, the improvement of RV precision in $H$ band is not seen in the comparison of $J$ and $H$ band in the results from~\citet{Reiners2010}. In addition, we report better RV precisions in $V$ band by a factor of 1.5.  
%The S/N we report in the table is the average S/N in a particular observational bandpass, and it is normalized at 1 $\mu m$ at spectral resolution of 60,000. In $V$ band, the maximum difference is 27\% after scaled with reported S/N. In $Y$ band, our results differ from theirs by 50\%. However, the comparison depends on the definition of S/N in their paper. in $Y$ band, average stellar flux in each spectral slice (with a wavelength range of 10 nm) can vary by a factor of 3. The results in $J$ band agree within 23\%. However, in $H$ band, our predicted RV precisions are a factor of 2 better than theirs. In further examination, we compare our results in $J$ and $H$ band and find that RV precisions in $H$ band are in general better than those in $J$ band. It is explained by the richer absorption features in $H$ band for an M9V star. On the contrast, the improvement of RV precision in $H$ band is not seen in the comparison of $J$ and $H$ band in the results from~\citet{Reiners2010} 

We have also conducted similar calculation to~\citet{Rodler2011} for an M9.5 dwarf ($T_{\rm{eff}}$=2200 K, $V\sin i$=5 $\rm{km}\cdot\rm{s}^{-1}$) and the comparison is presented in Table \ref{tab:Comp_Rodler}. Instead of finding $Y$ band, we find $K$ band gives the highest RV precision in telluric-contamination-free case, while $H$ band gives the highest RV precision in the case where telluric lines with an absorption depth more than 3\% are masked out in RV calculation. We also notice that our predicted RV precisions are less sensitive to spectral resolution. Note that stellar absorption lines typically become resolved by spectrograph after spectral resolution goes over 50,000, we do not expect RV precision increases steeply as spectral resolution goes well beyond 50,000.

\subsection{Current Precision vs. Signal of an Earth-like Planet in Habitable Zone}
\label{sec:HZ}

One of the most intriguing tasks in exoplanet science is to search and characterize Earth-like planets in the HZ. Over the past two decades, great advances have been seen but we have not yet  discovered another Earth. We are trying to answer several questions in the following discussion: 1, is it possible to detect an Earth-like planet in the HZ using the RV technique? 2, If so, at what S/N in which observational bandpass and for which spectral type? 3, Based on the current available RV calibration sources and knowledge of stellar noise, is it practical to detect an Earth-like planet in the HZ in the most optimistic case?

\subsubsection{Stellar Spectral Quality}
\label{sec:HZ_star}

\begin{figure}
\begin{center}
\includegraphics[width=16cm,height=12cm,angle=0]{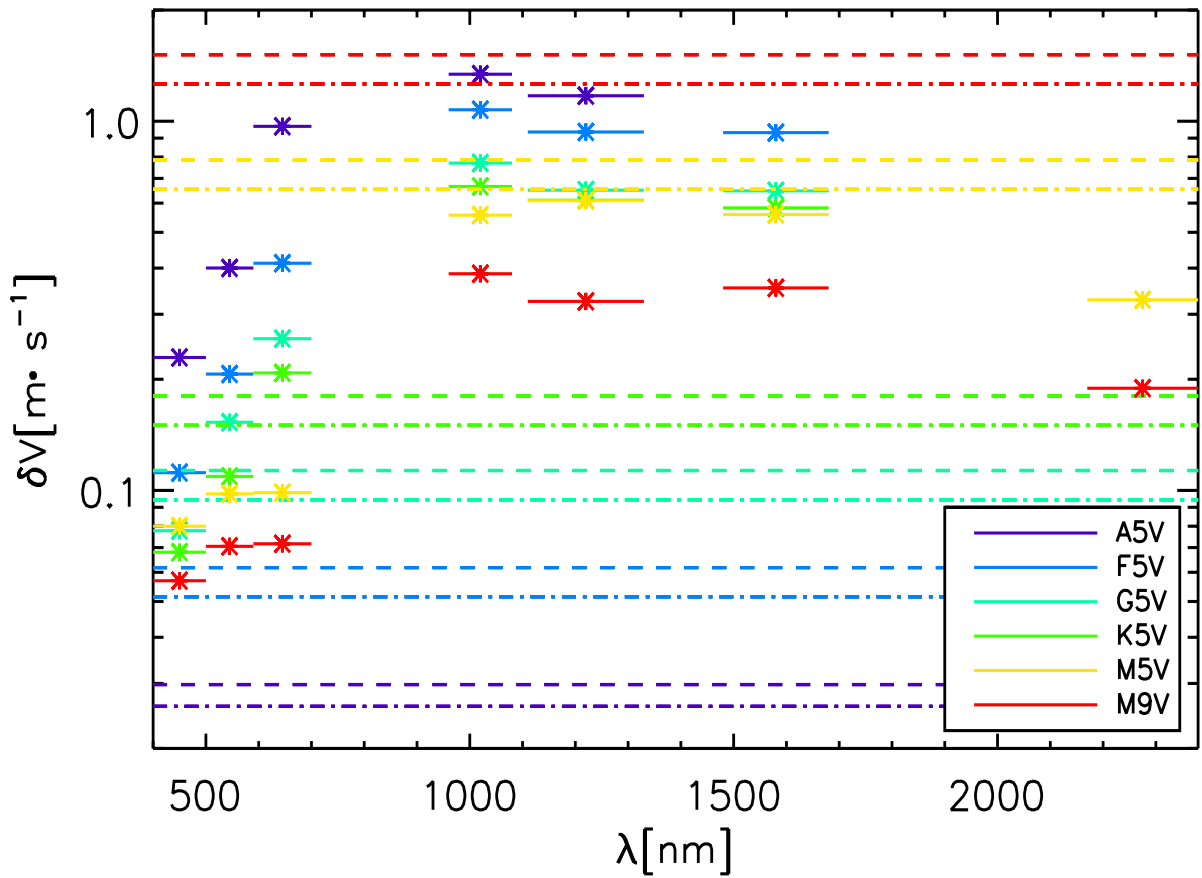}
\caption{RV precisions ($R$=120,000) considering spectral quality factor at a S/N of 425 as a function of observational bandpass for different spectral types from M9V to A5V (color-coded). Overplotted are RV signals of an Earth-like planet in the inner (dashed) and outer edge (dash-dotted) of the HZ. 
\label{fig:RV_wav_HZ_120000}}
\end{center}
\end{figure}

We first consider an ideal situation in which the RV precision is only determined by the Q factor of a stellar spectrum. The highest S/N per pixel obtainable for a single exposure is ~425 (assuming 30,000 ADU and a gain of 6 electron/ADU). Fig. \ref{fig:RV_wav_HZ_120000} shows the RV precisions obtainable at spectral resolution of 120,000 for different spectral types in different observational bandpasses. Overplotted are RV signals of an Earth-like planet in the inner (dashed) and outer edge (dash-dotted) of the HZ of a star with a certain spectral type (color coded). The position of the HZ is calculated based on ~\citet{Kasting1993}. The position of the HZ gets closer to the host star as stellar temperature and luminosity drops. The RV signal is enhanced by both the decreasing distance to the star and the decreasing stellar mass. Earth-like planet is detectable in every observational bandpass at a S/N as high as 425 for M dwarfs. $B$ and $V$ band bear the highest probability for K stars and $B$ band is the sweet sopt for G stars. Predicted RV precisions are not adequate to detect the signal of an Earth-like planet in the HZ around F and A stars with single exposure on a current typical CCD with 16-bit dynamic range. Table \ref{tab:SNR_SP} summarizes the S/N required for detection of an Earth-like planet in the HZ as a function of spectral type, in which we assume that a detection is possible when the RV precision is equal to the signal. Even though it only require a S/N of 17 for an M9V star in $B$ band to detect an habitable Earth-like planet, the exposure time could be as long as 1 hour even at the Keck telescope for a J=6 M9V star and there is no such bright late M type star in the sky. In addition, only 25 M stars are available with $J$ band magnitude less than 6~\citep{Lepine2005}. In comparison, $\sim$1 min exposure time at Keck will obtain a S/N of 175 for a B=8 star, which is adequate for detecting habitable Earth-like planet around a K5V star. 10\% instrument throughput is assumed in the above calculations. 

\subsubsection{Stellar Spectral Quality + RV Calibration Source + Atmosphere}
\label{sec:HZ_Star_RVcal}

\begin{figure}
\begin{center}
\includegraphics[width=16cm,height=12cm,angle=0]{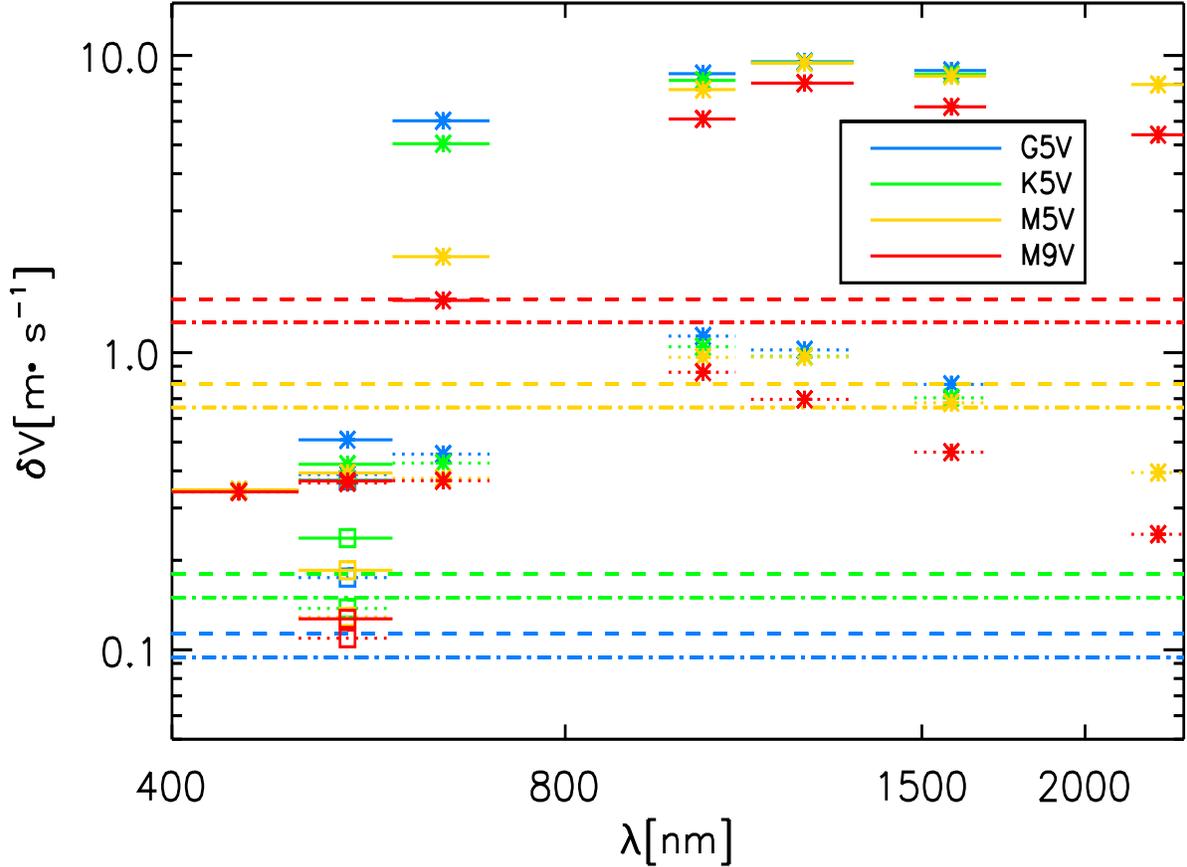}
\caption{RV precision ($R$=120,000) considering spectral quality factor (S/N=425), RV calibration uncertainties and telluric contamination as a function of observational bandpass for different spectral types from M9V to G5V (color-coded). Overplotted are RV signals of an Earth-like planet in the inner (dashed) and outer edge (dash-dotted) of the HZ. Solid lines represent non-telluric-removal cases while dotted lines represent cases in which 99.9\% of the strength of telluric lines is removed. Squares in $V$ band represent Iodine cell method and asterisks in $V$ band represent Th-Ar lamp method. Refer to Table \ref{tab:RV_cal} for RV calibration sources in different observational bandpasses.
\label{fig:RV_wav_HZ_atm_120000}}
\end{center}
\end{figure}

Current RV precision is not only restricted by the photon-limited uncertainty determined by a stellar spectrum, but also by the uncertainties brought by an RV calibration source and the telluric contamination from the Earth's atmosphere. Fig. \ref{fig:RV_wav_HZ_atm_120000} shows the RV precisions taking into consideration of Q factors, RV calibration uncertainties and telluric contamination. Bracketing calibration (at a S/N of 425) is considered in the calculations in which the S/N of calibration is not determined by the continuum of the observed star. Two cases are discussed for telluric contamination, in one case no telluric removal is attempted (solid) while in the other case 99.9\%  of telluric line strength is removed (dotted). For the visible bands (i.e., $B$, $V$ and $R$ band), RV precision in $B$ band is barely affected by telluric contamination (10 $\rm{m}\cdot\rm{s}^{-1}$ random RV of telluric lines is assumed in the calculations) but limited by the RV calibration uncertainty due to a Th-Ar lamp. There are two RV calibration sources considered in the $V$ band, a Th-Ar lamp and a Iodine absorption cell, the latter one provides higher calibration precision in the Bracketing case. 

Even though only 2.4\% of the wavelength range is affected by telluric line contamination (\S \ref{sec:Comp_Prev}), handling telluric lines is still very important. The RV uncertainty budget of a K5V star in Table \ref{tab:atm_example} shows an example in which the RV precision is worse than detection limit if no telluric line removal is involved while it is below the detection limit in the 99.9\% removal case ($\alpha$=0.001). This example address the importance of telluric line removal in the search of an Earth-like planet even in the visible band where telluric contamination is less severe than the NIR band. After 99.9\% of telluric line strength is removed, the RV uncertainty of a K5V star is dominated by the spectral quality of a K5V star and the RV calibration source.

In comparison, in the NIR ($Y$, $J$, $H$ and $K$ band), the RV uncertainties are dominated by telluric contamination, resulting RV precisions at 5$-$10 $\rm{m}\cdot\rm{s}^{-1}$ that are not adequate in habitable Earth-like planet detections. A similar example of how telluric contamination raises the floor of RV uncertainty is also given for an M5V star in $K$ band in Table \ref{tab:atm_example}. In the 99.9\% removal case, RV uncertainties in the NIR are no longer mainly dominated by telluric contamination, but by spectral quality factor. To sum up, telluric line removal is an important and indispensable step toward the discovery of an Earth-like planet even in the visible band. After telluric lines are successfully removed from observed stellar spectrum, the RV precision is limited by the uncertainty caused by stellar spectral quality and RV calibration sources. 

After completely removing the telluric contamination, we compare our prediction of RV uncertainties and what is reported from HARPS instrument~\citep{Mayor2003}. An example of HD 47186~\citep{Bouchy2009} is given in Table \ref{tab:harps_example}. HD 47186 is a G5V star with a $V\sin i$ of 2.2 $\rm{km}\cdot\rm{s}^{-1}$, the best achievable RV precision for this star is 0.3 $\rm{m}\cdot\rm{s}^{-1}$ at a S/N of 250 according to~\citet{Bouchy2009}. Our prediction indicates that an RV precision of 0.24 $\rm{m}\cdot\rm{s}^{-1}$ is possible to achieve at the same S/N for the same wavelength coverage. The difference may come from those uncounted factors in our calculation, for example, stellar noise. However, our prediction of RV precision is within 20\% to precision from real observation.

\subsubsection{Stellar Spectral Quality + RV Calibration Source + Stellar Noise}
\label{sec:OptiBand_Star_RVcal_Noise}

\begin{figure}
\begin{center}
\includegraphics[width=16cm,height=12cm,angle=0]{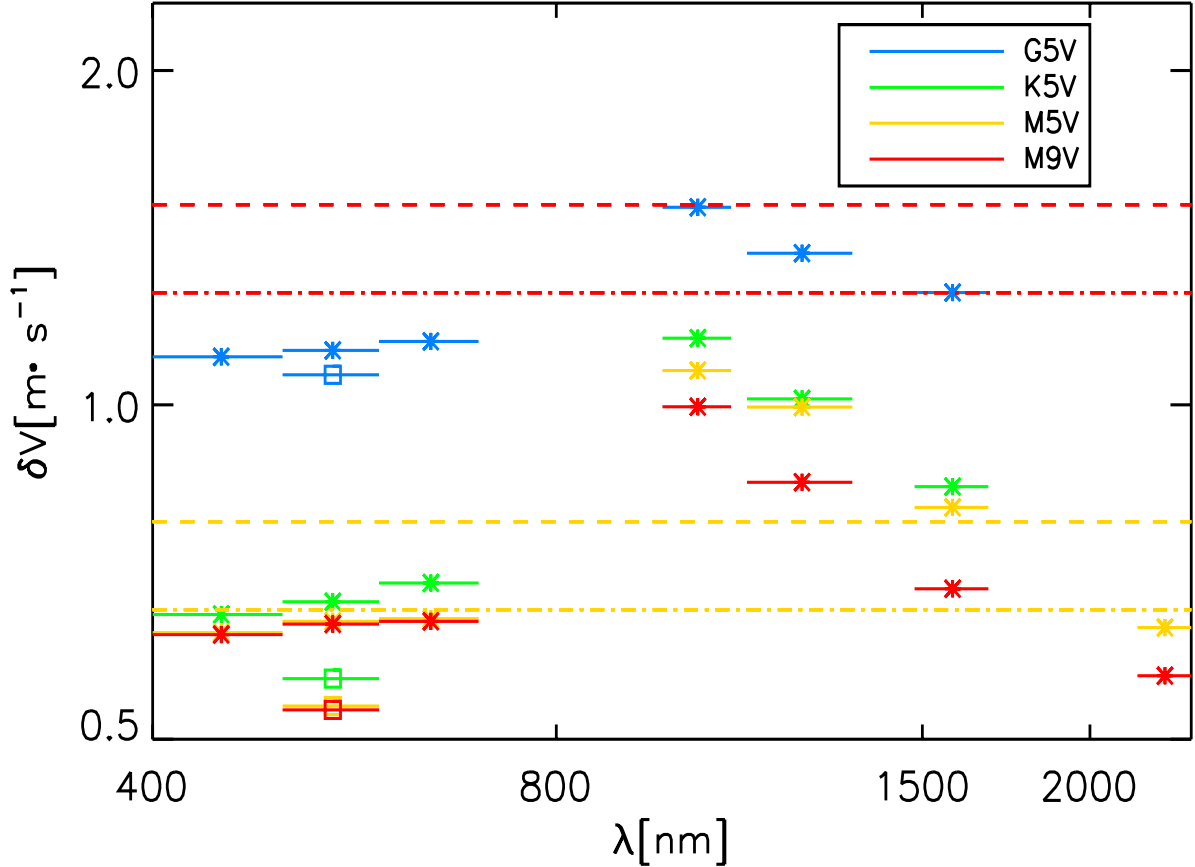}
\caption{RV precision ($R$=120,000) considering spectral quality factor (S/N=425), RV calibration uncertainties and stellar noise as a function of observational bandpass for different spectral types from M9V to G5V (color-coded). Overplotted are RV signals of an Earth-like planet in the inner (dashed) and outer edge (dash-dotted) of the HZ. HZs of G and K type stars are not plotted because they are out of reach based on the predicted RV precision. Squares in $V$ band represent Iodine cell method and asterisks in $V$ band represent Th-Ar lamp method. Refer to Table \ref{tab:RV_cal} for RV calibration sources in different observational bandpasses.
\label{fig:RV_wav_HZ_sn_120000}}
\end{center}
\end{figure}

Assuming the telluric lines are perfectly measured and removed, we consider the obtainable RV precision based on stellar spectral quality, RV calibration precision and stellar noise. Stellar noise of different spectral type is estimated in \S \ref{sec:StellarNoise} based on~\citet{Dumusque2011}. Fig. \ref{fig:RV_wav_HZ_sn_120000} shows predicted RV precision, it is clear that the RV precision for G and K type stars is not adequate for detecting Earth-like planet in the HZ after stellar noise is taken into consideration. However, it is still possible to detect Earth-like planets around M dwarfs because of relatively larger RV signal. For an M5V star, visible band and K band provide adequate precision for an Earth-like planet detection while all bandpasses allow an Earth-like planet detection for an M9V star.  

In order to compare our predicted RV precision to observations, we choose 69 planets detected by HARPS since 2004 after an instrument upgrade~\citep{Mayor2003} and plot the RMS errors of Keplerian orbit fitting as a function $T_{\rm{eff}}$. The minima of three subsets (corresponding to G, K and M stype stars) are found based on $T_{\rm{eff}}$. In comparison, the predicted RV predictions (after combining results from $B$, $V$ and $R$ bands) for a G5V, K5V and M5V star are plotted as open diamonds. Since the predicted RV precision is based on an optimistic case, we compare the predictions with the RMS minima we find in the observation. For M dwarfs, the minimum of RMS errors is found at 0.8 $\rm{m}\cdot\rm{s}^{-1}$ with GJ 674 b~\citep{Bonfils2007}. In comparison, our prediction is 0.62 $\rm{m}\cdot\rm{s}^{-1}$ considering stellar spectral quality factor, RV calibration error and stellar noise. If $\sim$0.5 $\rm{m}\cdot\rm{s}^{-1}$ instrumental uncertainty as mentioned in ~\citet{Bonfils2007} is added in quadrature, our prediction is well matched with HARPS M dwarfs observation in the best case scenario.~\citet{Lovis2006} reported 0.64 $\rm{m}\cdot\rm{s}^{-1}$ RMS errors for a planet system of a K0V star (i.e., HD 69830), which is consistent with our prediction of 0.65 $\rm{m}\cdot\rm{s}^{-1}$. It is plotted in the bin with $T_{\rm{eff}}$ between 5000 K and 6000 K because reported $T_{\rm{eff}}$ of 5385 K. For G type stars,~\citet{Bouchy2009} reported 0.91 $\rm{m}\cdot\rm{s}^{-1}$ RMS error for HD 47186 b and c, a planetary system around a G5V star. In comparison, we predict a total RV uncertainty of 1.1 $\rm{m}\cdot\rm{s}^{-1}$. The overestimation of RV measurement uncertainty is possibly due to an overestimation of stellar noise or an increasing S/N because of multiple measurements in real observation. 

We predict a total RV measurement uncertainty of 0.62, 0.65 and 1.1 $\rm{m}\cdot\rm{s}^{-1}$ for spectral type M5V, K5V and G5V considering stellar spectral quality, RV calibration and stellar noise. According to the calculations in \S \ref{sec:StellarNoise}, RV uncertainty due to stellar noise is 0.52, 0.55 and 1.05 $\rm{m}\cdot\rm{s}^{-1}$ for the above three types of stars, accounting for 70.3\%, 71.6\% and 91.1\% of total RV measurement uncertainty. Based on comparisons of our predictions and observation, we therefore conclude that stellar noise is one major contributor in error budget of precision Doppler measurement. M dwarfs should be the primary targets in search of Earth-like planets in the HZ. Unlike G and K stars, the RV signal of Earth-like planets in the HZ is not overwhelmed by stellar noise for M dwarfs in the most optimistic case. 

\section{Summary and Discussion}
\label{sec:Conclusion}

We provide a method of practically estimating the photon-limited RV precision based on the spectral quality factor, stellar rotation, RV calibration uncertainty, stellar noise and telluric line contamination. The methodology described and the results presented in this paper can be used for design and optimization of planned and ongoing precision Doppler planet surveys. For pure consideration of stellar spectral quality without artificial rotationally broadening the absorption line profile, the optimal band for RV planet search is $B$ band for a wide range of spectral types from K to A, while it is $R$ or $K$ band for mid-late type M dwarfs. Nevertheless, the above conclusion remains unchanged after considering typical stellar rotation of each spectral type. However, F and A stars become unsuitable for precision RV measurements because of typically fast stellar rotation. We confirm the finding in~\citet{Reiners2010} that the NIR Doppler measurements gain advantage for mid-late M dwarfs. However, instead of finding $Y$ band as the optimal band considering stellar spectrum quality and telluric masking, we find that both $Y$ and $H$ bands give the highest RV precision among $V$, $Y$, $J$ and $H$ bands. In a comparison to~\citet{Rodler2011}, we find $K$ band is the optimal band for precision Doppler measurement in a telluric-free case and $H$ band is optimal in a telluric-masking case, while they found $Y$ band gives the highest RV precision in both cases. Fundamental photon-limited RV precision for evolved stars has been discussed by~\citet{Jiang2011}, which is valuable for ongoing RV planet search around retired stars discussed in~\citet{Johnson2007}. 

\begin{figure}
\begin{center}
\includegraphics[width=16cm,height=12cm,angle=0]{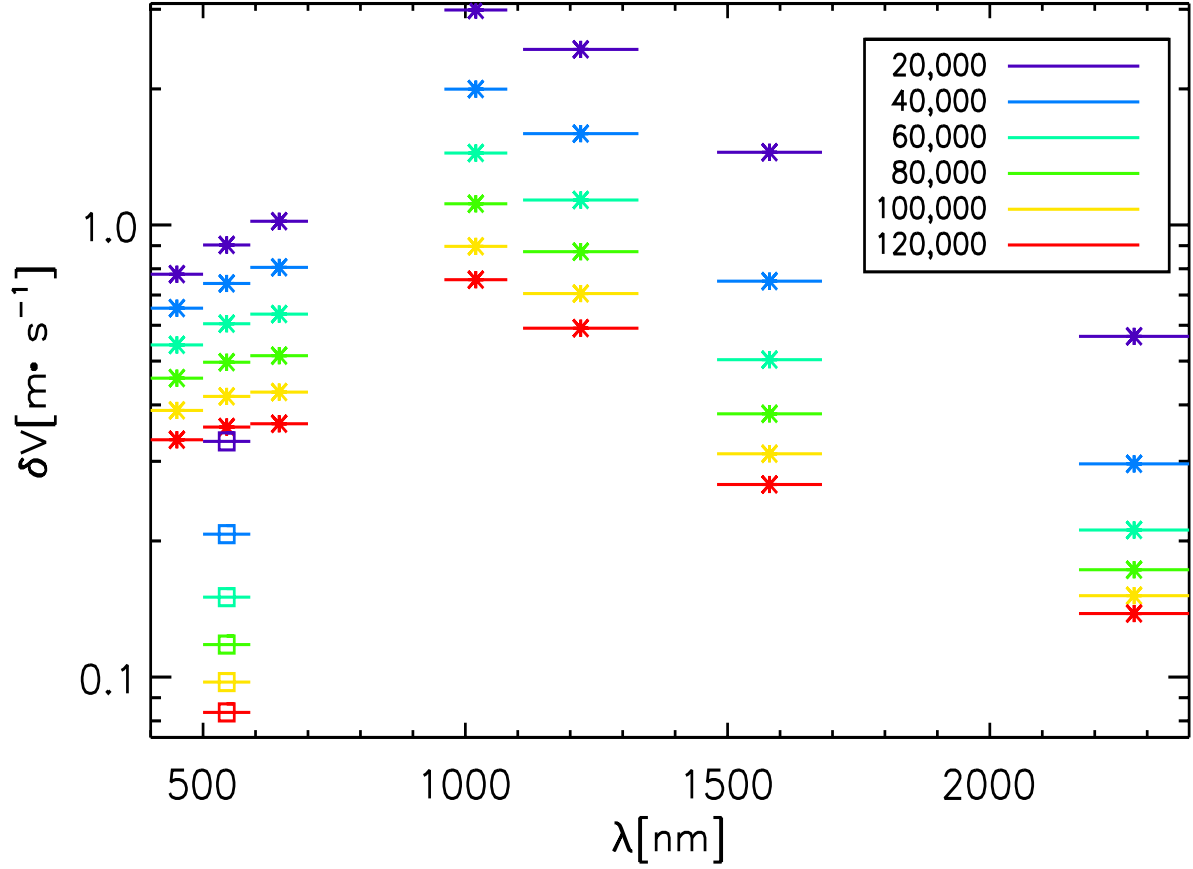}
\caption{RV calibration uncertainties as a function of observational bandpass at different spectral resolutions (color-coded). Squares in $V$ band represent Iodine cell method and asterisks in $V$ band represent Th-Ar lamp method. Refer to Table \ref{tab:RV_cal} for RV calibration sources in different observational bandpasses.
\label{fig:Un_Cal_Res}}
\end{center}
\end{figure}

We also consider the uncertainties brought by current available RV calibration sources at different spectral resolutions (Fig. \ref{fig:Un_Cal_Res}). Sub $\rm{m}\cdot\rm{s}^{-1}$ calibration precision can be reached for each observational bandpass. Note that the Q factors may change as gas pressure, length of light path and temperature changes. The precision also depends on the methods used in the RV calibration. We categorized the calibration methods into several cases: Superimposing, in which the calibration spectrum is imprinted onto a stellar spectrum; Non-Common Path and Bracketing, in which the calibration is conducted either spatially or temporally. The former method depends on stellar flux while the latter one can only be applicable for very stable instruments. There are other calibration sources we have not included into the discussions in this study, for example, laser combs~\citep{Steinmetz2008,Li2008}, the Fabry-Perot calibrator~\citep{Wildi2010} and the Monolithic Michelson Interferometer~\citep{Wan2010}. Once they become more economically affordable or more technically ready, the RV precision will be greatly improved in the future. 

For the first time we have quantitatively estimated the uncertainty caused by the residual of telluric contamination removal for high resolution echelle spectroscopy method. Depending on the telluric absorption, different observational bandpasses are affected differently. $B$ band is the least sensitive to telluric contamination because there are barely any telluric absorption features in $B$ band. However, the NIR bands suffer the most in precision RV measurements because the stellar absorption lines and telluric lines are mixed together severely in this spectral region. Only when $\alpha\leq$0.01, i.e., more than 99\% of strength of telluric lines is removed, the advantage of NIR observation of mid-late type M dwarfs begins to show, which is a factor of 3 improvement. This quantitative method in estimating the RV uncertainty induced by telluric contamination can be easily adapted to other problems, for example, estimating the moon light contamination.

Besides telluric line removal, telluric line masking has also been discussed in several of previous studies~\citep{Reiners2010, Rodler2011, Wang2011}. In ~\citet{Reiners2010}, telluric absorption with depth more than 2\% and 30 $\rm{km}\cdot\rm{s}^{-1}$ in the vicinity is blocked out when measuring RV. Based on this blocking criterium, the photon-limited RV uncertainty, $\delta v_{rms, S}$(refer to Equation (\ref{eq:simple_example_tulleric_2})) , for an M9V star at $\rm{R}$=100,000 is 3.9, 2.2, 3.9, 2.2 $\rm{m}\cdot\rm{s}^{-1}$ in $V$, $Y$, $J$ and $H$ band respectively (see Table \ref{tab:Comp_Reiners}). In comprison, $\delta v_{rms, N}$ is 71.3, 5.8, 6.5, 3.7 $\rm{m}\cdot\rm{s}^{-1}$ in $V$, $Y$, $J$ and $H$ band respectively. Except for $V$ band, $\delta v_{rms, S}$ and $\delta v_{rms, N}$ are at the same order of magnitude, and the uncertainty caused by telluric absorption cannot be neglected even though that the spectral region with any telluric absorption of more than 2\% is blocked. If more strict criterium of telluric line masking is applied, fewer photons are considered in measuring the RV, which effectively increases the photon-limited RV uncertainty. In order to reach photon-limited RV precision predicted by pure consideration of spectral Q factor, telluric removal should be applied in which telluric contamination is measured or modeled and then removed from measured stellar spectrum.

RV uncertainty due to stellar granulation is taken into consideration in this paper. High frequency ($\sim$min) stellar noise such as p-mode oscillations usually have a RV amplitude of 0.1 to 4.0 $\rm{m}\cdot\rm{s}^{-1}$ ~\citep{Schrijver2000} and they can be averaged out within typical 10$-$15 exposure time. RV uncertainties due to low frequency ($10-100$ day) stellar noise such as stellar spots have been discussed in recent papers, for example, ~\citet{Desort2007} and ~\citet{Reiners2010}. The amplitudes of spot-induced RV range from one to several hundred $\rm{m}\cdot\rm{s}^{-1}$. Since stellar spot induced RV uncertainties are periodic and therefore can be modeled and removed, however, the amplitude of residual is unknown at this stage. 

We compare the RV precision based on stellar spectral quality and the signal of an Earth-like planet in the HZ of a star with a certain spectral type. We find that it is likely to detect a habitable Earth-like planet around G,K and M stars while it is too demanding to detect one around F and A stars. $B$ band is the optimal band for G and K stars and $K$ band for M dwarfs. After considering practical issues such as telluric contamination, we find that, except for $B$ band, every observational bandpass is affected by telluric contamination to some extent. The major RV measurement uncertainty comes from telluric contamination, which overwhelms the RV signal of an habitable Earth-like planet around G and K stars. Surprisingly, telluric contamination becomes an issue in $V$ band even there is only 2.4\% of spectral region affected by telluric lines. After telluric lines are removed at a very high level, i.e., $\alpha\leq$0.001, the error from RV calibration becomes the major contributor of Doppler measurement uncertainty. After stellar noise (granulation only) is taken into consideration, which is dominant contributor to RV uncertainty, M dwarfs become the only type of star that is suitable for the search for Earth-like planets in the HZ. 

The RV precision in the discussion of habitable Earth detectability considers four factors: stellar spectral quality, RV calibration uncertainty, stellar noise and telluric contamination. However, the discussion of stellar noise should be treated with great caution for several reasons: 1, stellar noise is not very well understood and characterized at this stage; 2, it is different from case to case and therefore it is difficult to draw a general conclusion; 3, a habitable planet search is different from a planet survey, the targets are chosen in favor of detection at the best case scenario, for example, high stellar flux, slow stellar rotation, low stellar activity and low stellar noise and so on. Therefore, the stellar noise assumed in this study is in the best case scenario according to current theory and observation. In addition, we assume the highest signal within linear range (30,000 ADU) for current typical CCD (16-bit dynamic range) in single exposure in the discussion, note that the S/N can also be improved by multiple independent measurements. The S/Ns required for Earth-like planet detections are provided in Table \ref{tab:SNR_SP} based on stellar spectral quality. Please note that the HZ changes over time as the luminosity of the host star changes. It also depends on properties of a planet such as atmosphere composition, albedo and orbit. The purpose of discussion in this paper is to provide a basic idea of the comparison of current best obtainable RV precision to a typical RV signal of an habitable Earth-like planet. 

We  acknowledge the support from NSF with grant NSF AST-0705139, NASA with
grant NNX07AP14G (Origins), UCF-UF SRI program, DoD ARO Cooperative Agreement W911NF-09-2-0017, 
Dharma Endowment Foundation and the University of Florida. We thank Stephen Redman for kindly providing us the lines list of a U-Ne emission lamp. We thank Suvrath Mahadevan, Cullen Blake, Peng Jiang and Mark Keremedjiev for valuable discussion and comments.

%\begingroup
%\scriptsize \setlength{\itemsep}{-2mm}
\bibliographystyle{apj}
\bibliography{mybib_JW_JG_V5}{}

%\endgroup

%\clearpage

%\begin{figure}
%\begin{center}
%\includegraphics[width=16cm,height=12cm,angle=0]{RV_wav_20000}
%\caption{
%\label{fig:RV_wav_20000}}
%\end{center}
%\end{figure}
%
%\begin{figure}
%\begin{center}
%\includegraphics[width=16cm,height=12cm,angle=0]{RV_wav_60000}
%\caption{
%\label{fig:RV_wav_60000}}
%\end{center}
%\end{figure}

\begin{figure}
\begin{center}
\includegraphics[width=16cm,height=12cm,angle=0]{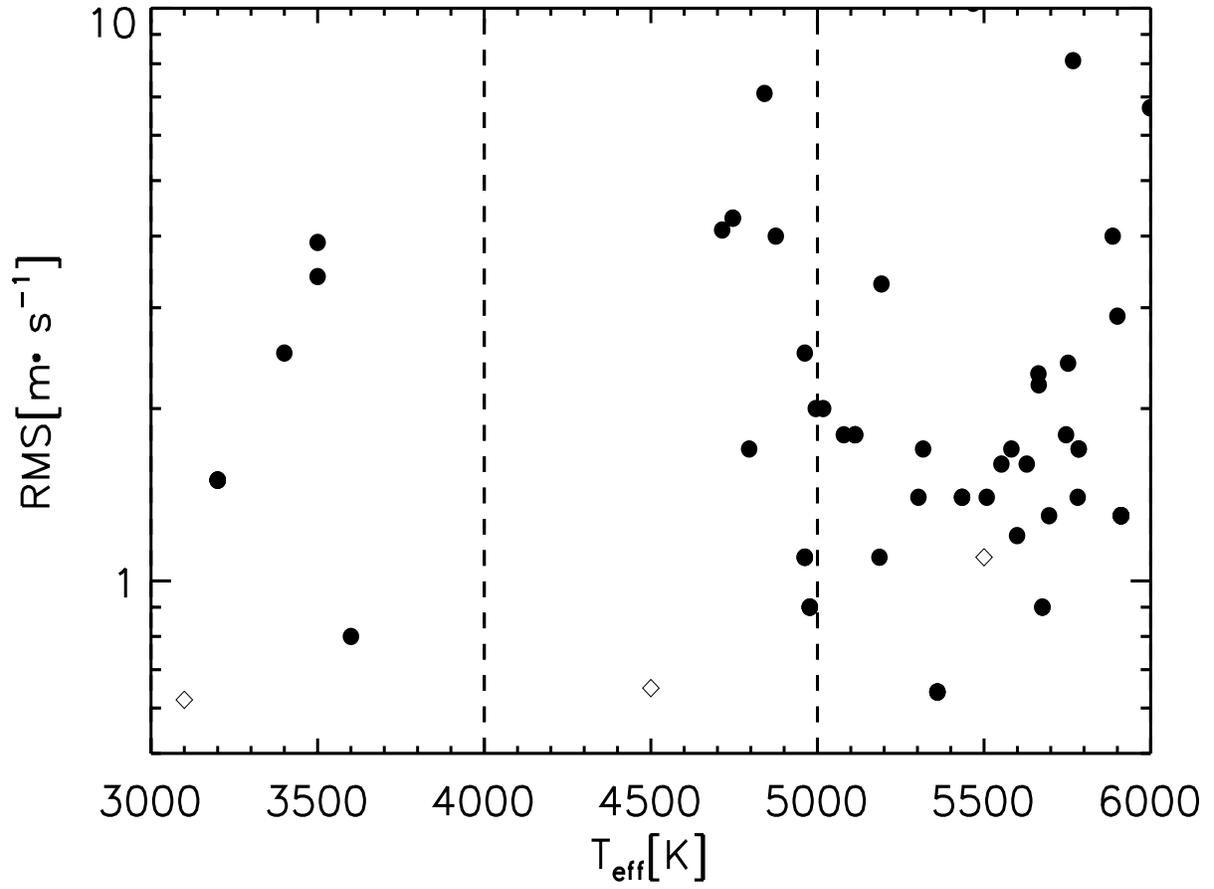}
\caption{RMS error of Keplerian orbit fitting for planets detected by HARPS since 2004. Predicted RV precisions considering stellar noise for different stellar types are overplotted as open diamonds.
\label{fig:rms_harps}}
\end{center}
\end{figure}

\clearpage

\begin{deluxetable}{ccc}

%\tabletypesize{\footnotesize}
\tablewidth{355pt}

\tablecaption{Definition of observational bandpasses used in this study: center wavelength and wavelength range}

\tablehead{\colhead{Band} & \colhead{$\lambda_0$} & \colhead{$\lambda_{min}-\lambda_{max}$} \\
\colhead{} & \colhead{(nm)} & \colhead{(nm)} }

%% All data must appear between the \startdata and \enddata commands
\startdata 
\label{tab:SpecBand}

$B$ & 450 & 400$-$500 \\
$V$ & 545 & 500$-$590 \\
$R$ & 660 & 590$-$730 \\
$Y$ & 1020 & 960$-$1080 \\
$J$ & 1220 & 1110$-$1330 \\
$H$ & 1580 & 1480$-$1680 \\
$K$ & 2275 & 2170$-$2380 \\
\enddata

\end{deluxetable}

\begin{deluxetable}{cccccccc}

\tabletypesize{\footnotesize}
\tablewidth{450pt}

\tablecaption{RV uncertainties caused by calibration sources at different spectral resolutions}

\tablehead{\colhead{$\rm{R}$} & \colhead{$B$}  & \colhead{$V$} & \colhead{$R$} & \colhead{$Y$} & \colhead{$J$} & \colhead{$H$} & \colhead{$K$} \\ \colhead{} & \colhead{Th-Ar$^a$} & \colhead{Th-Ar, Iodine$^b$} & \colhead{Th-Ar} & \colhead{U-Ne$^c$} & \colhead{U-Ne} & \colhead{Mixed cell$^d$} & \colhead{Ammonia$^e$} \\ \colhead{} & \colhead{($\rm{m}\cdot\rm{s}^{-1}$)} & \colhead{($\rm{m}\cdot\rm{s}^{-1}$)} & \colhead{($\rm{m}\cdot\rm{s}^{-1}$)} & \colhead{($\rm{m}\cdot\rm{s}^{-1}$)} & \colhead{($\rm{m}\cdot\rm{s}^{-1}$)} & \colhead{($\rm{m}\cdot\rm{s}^{-1}$)} & \colhead{($\rm{m}\cdot\rm{s}^{-1}$)}     }

%% All data must appear between the \startdata and \enddata commands
\startdata 
\label{tab:RV_cal}
20,000 & 0.78 & 0.90, 0.33 & 1.0 & 3.0 & 2.4 & 1.4 & 0.57 \\
40,000 & 0.65 & 0.74, 0.21 & 0.81 & 2.0 & 1.6 & 0.75 & 0.30 \\
60,000 & 0.54 & 0.60, 0.15 & 0.63 & 1.4 & 1.1 & 0.50 & 0.21 \\
80,000 & 0.46 & 0.50, 0.12 & 0.51 & 1.1 & 0.87 & 0.38 & 0.17 \\
100,000 & 0.38 & 0.41, 0.10 & 0.43 & 0.90 & 0.70 & 0.31 & 0.15 \\
120,000 & 0.33 & 0.36, 0.08 & 0.36 & 0.76 & 0.59 & 0.27 & 0.14 \\

\enddata

\tablecomments{a:~\citet{Lovis2007}; b:~\citet{Butler1996}; 
c:~\citet{Redman2011}; d:~\citet{Mahadevan2009}; e:~\citet{Bean2010}.
}

\end{deluxetable}

\begin{deluxetable}{cccccccc}

\tabletypesize{\footnotesize}
\tablewidth{450pt}

\tablecaption{Photon-limited RV uncertainties based on stellar spectral quality at different spectral resolutions for different spectral types, average S/N per pixel is reported in perentheses}

\tablehead{\colhead{Spec. Type} & \colhead{$B$}  & \colhead{$V$} & \colhead{$R$} & \colhead{$Y$} & \colhead{$J$} & \colhead{$H$} & \colhead{$K$} \\ \colhead{} & \colhead{($\rm{m}\cdot\rm{s}^{-1}$)} & \colhead{($\rm{m}\cdot\rm{s}^{-1}$)} & \colhead{($\rm{m}\cdot\rm{s}^{-1}$)} & \colhead{($\rm{m}\cdot\rm{s}^{-1}$)} & \colhead{($\rm{m}\cdot\rm{s}^{-1}$)} & \colhead{($\rm{m}\cdot\rm{s}^{-1}$)} & \colhead{($\rm{m}\cdot\rm{s}^{-1}$)}     }

%% All data must appear between the \startdata and \enddata commands
\startdata 
\label{tab:RV_spec}
\\
\hline
$\rm{R}$=20,000 \\
\hline
A5V&4.7(211.7)&8.3(209.0)&21.5(198.3)&23.9(173.2)&22.6(155.1)&62.9(126.8)&...(...)\\
F5V&3.8(166.9)&6.3(177.1)&12.8(182.1)&23.3(173.2)&20.9(164.2)&27.2(148.9)&...(...)\\
G5V&3.3(126.4)&5.1(145.6)&8.6(163.3)&20.8(173.2)&15.9(171.9)&14.9(168.0)&...(...)\\
K5V&3.7(88.8)&4.8(110.9)&7.5(141.0)&17.3(173.2)&13.5(181.8)&11.2(199.8)&...(...)\\
M5V&14.7(26.8)&11.7(44.0)&9.0(66.2)&12.8(173.2)&12.0(203.0)&9.5(205.6)&6.8(191.6)\\
M9V&28.2(9.3)&18.4(17.9)&13.8(26.6)&8.2(173.2)&4.6(250.5)&5.4(246.3)&2.7(250.9)\\
\hline
$\rm{R}$=60,000 \\
\hline
A5V&1.5(122.2)&2.7(120.7)&6.9(114.5)&10.0(100.0)&9.6(89.5)&24.6(73.2)&...(...)\\
F5V&1.1(96.4)&1.8(102.3)&3.6(105.1)&8.8(100.0)&7.9(94.8)&9.4(86.0)&...(...)\\
G5V&1.0(73.0)&1.6(84.0)&2.5(94.3)&6.9(100.0)&5.7(99.3)&5.6(97.0)&...(...)\\
K5V&1.2(51.3)&1.5(64.0)&2.3(81.4)&5.9(100.0)&5.1(105.0)&4.6(115.3)&...(...)\\
M5V&4.6(15.5)&3.5(25.4)&2.5(38.2)&4.8(100.0)&4.4(117.2)&3.8(118.7)&2.4(110.6)\\
M9V&9.3(5.4)&5.9(10.4)&4.2(15.3)&3.3(100.0)&1.8(144.6)&2.0(142.2)&1.0(144.9)\\
\hline
$\rm{R}$=120,000 \\
\hline
A5V& 1.1(86.4)& 2.0(85.3)& 5.1(81.0)& 8.0(70.7)& 7.9(63.3)& 18.9(51.8)&...(...)\\
F5V& 0.7(68.2)& 1.2(72.3)& 2.4(74.3)& 6.4(70.7)& 5.9(67.0)& 6.5(60.8)&...(...)\\
G5V& 0.6(51.6)& 1.1(59.4)& 1.6(66.7)& 4.6(70.7)& 3.9(70.2)& 4.0(68.6)&...(...)\\
K5V& 0.8(36.3)& 1.0(45.3)& 1.5(57.6)& 4.0(70.7)& 3.5(74.2)& 3.0(81.6)&...(...)\\
M5V& 3.1(11.0)& 2.3(18.0)& 1.5(27.0)& 3.3(70.7)& 3.1(82.9)& 2.8(83.9)&1.8(78.2)\\
M9V& 6.4(3.8)& 4.1(7.3)& 2.8(10.8)& 2.3(70.7)& 1.3(102.3)& 1.5(100.6)&0.8(102.4)\\

\enddata

\end{deluxetable}

\begin{deluxetable}{ccc}

%\tabletypesize{\footnotesize}
\tablewidth{400pt}

\tablecaption{Spectral Type, corresponding $T_{\rm{eff}}$, and typical stellar rotation $V\sin i$}

\tablehead{\colhead{Spectral Type} & \colhead{$T_{\rm{eff}}$} & \colhead{V$\sin i$}  \\
\colhead{} & \colhead{(K)} & \colhead{($\rm{km}\cdot\rm{s}^{-1}$)}  }

%% All data must appear between the \startdata and \enddata commands
\startdata 
\label{tab:TypeTeff}

A0V & 9600 & 131.0 \\ 
A2V & 9000 & 108.0 \\ 
A5V & 8400 & 85.5 \\ 
A8V & 7800 & 62.5 \\ 
F0V & 7400 & 47.5 \\ 
F2V & 7000 & 32.0 \\ 
F5V & 6750 & 23.0 \\ 
F8V & 6250 & 6.5 \\
G0V & 6000 & 3.5 \\
G2V & 5750 & 2.2 \\
G5V & 5500 & 1.7 \\
G8V & 5250 & 1.8 \\
K0V & 5000 & 1.9 \\
K2V & 4750 & 1.8 \\
K5V & 4500 & 2.0 \\
K8V & 4000 & 2.5 \\
M0V & 3750 & 2.8 \\
M2V & 3500 & 2.8 \\
M5V & 3100 & 3.9 \\
M8V & 2600 & 6.8 \\
M9V & 2400 & 8.0 \\
\enddata

\end{deluxetable}

\begin{deluxetable}{ccccccc}

%\tabletypesize{\footnotesize}
\tablewidth{500pt}

\tablecaption{Comparison of Q factors from our results to ~\citet{Bouchy2001}}

\tablehead{\colhead{$T_{\rm{eff}}$} & \colhead{log$g$} & \colhead{$V_{t}$} & \colhead{} & \colhead{V$\sin i$} & \colhead{} & \colhead{}  \\  \colhead{(K)} & \colhead{($\rm{cm}\cdot\rm{s}^{-1}$)} & \colhead{($\rm{km}\cdot\rm{s}^{-1}$)} & \colhead{0 $\rm{km}\cdot\rm{s}^{-1}$} & \colhead{4 $\rm{km}\cdot\rm{s}^{-1}$} & \colhead{8 $\rm{km}\cdot\rm{s}^{-1}$} & \colhead{12 $\rm{km}\cdot\rm{s}^{-1}$} }

%% All data must appear between the \startdata and \enddata commands
\startdata 
\label{tab:Comp_Bouchy}

4500 & 4.5 &1.0 & 30238(34940) & 17235(17080) & 8700(8440) &  5793(5380) \\ 
5000 & 4.5 &1.0 & 30001(33405) & 16607(16140) & 8305(7815) &  5432(4930) \\ 
5500 & 4.5 &1.0 & 26892(30375) & 14858(14700) & 7397(7020) &  4700(4385) \\ 
\enddata

\end{deluxetable}

%\begin{deluxetable}{ccccccccc}
%
%%\tabletypesize{\footnotesize}
%\tablewidth{400pt}
%
%\tablecaption{Comparison of our results to ~\citet{Reiners2010}}
%
%\tablehead{\colhead{$\rm{R}$} & \multicolumn{4}{c}{S/N} & \multicolumn{4}{c}{$\delta v$($\rm{m}\cdot\rm{s}^{-1}$)}  \\
%\colhead{} & \colhead{$V$} & \colhead{$Y$} & \colhead{$J$} & \colhead{$H$} & \colhead{$V$} & \colhead{$Y$} & \colhead{$J$} & \colhead{$H$}  }
%
%%% All data must appear between the \startdata and \enddata commands
%\startdata 
%\label{tab:Comp_Reiners}
%& & & & & & & & \\
%\hline
%\multicolumn{9}{l}{Results from this study for M9V ($T_{\rm{eff}}=2400 K$)} \\
%\hline
%60000 & 14 & 126 & 149 & 128 & 5.4 & 2.9 & 4.5 & 2.0 \\
%80000 & 12 & 108 & 129 & 112 & 4.7 & 2.4 & 3.8 & 1.8 \\
%100000 & 11 & 97 & 116 & 100 & 4.2 & 2.2 & 3.4 & 1.6 \\
%\hline
% \multicolumn{9}{l}{Results from ~\citet{Reiners2010} for M9V} \\
%\hline
%60000 & 12 & 100 & 134 & 128 & 8.0 & 2.2 & 4.6 & 4.0 \\
%80000 & 10 & 86 & 116 & 111 & 6.2 & 1.7 & 3.5 & 3.5 \\
%100000 & 9 & 77 & 104 & 99 & 5.3 & 1.5 & 2.9 & 3.3 \\
%
%\enddata
%
%
%\end{deluxetable}

\begin{deluxetable}{ccccccccccccc}

%\tabletypesize{\footnotesize}
\tablewidth{400pt}

\tablecaption{Comparison of predicted RV precision (in the unit of $\rm{m}\cdot\rm{s}^{-1}$) between our results to ~\citet{Reiners2010} for an M9 dwarf ($T_{\rm{eff}}$=2400 K, $V\sin i$=0 $\rm{km}\cdot\rm{s}^{-1}$)}

\tablehead{\colhead{$\rm{R}$} & \multicolumn{4}{c}{S/N} & \multicolumn{4}{c}{This study} & \multicolumn{4}{c}{~\citet{Reiners2010}} \\
\colhead{} & \colhead{$V$} & \colhead{$Y$} & \colhead{$J$} & \colhead{$H$} & \colhead{$V$} & \colhead{$Y$} & \colhead{$J$} & \colhead{$H$} & \colhead{$V$} & \colhead{$Y$} & \colhead{$J$} & \colhead{$H$} }

%% All data must appear between the \startdata and \enddata commands
\startdata 
\label{tab:Comp_Reiners}
60000 & 12 & 100 & 134 & 128 & 5.1 & 3.0 & 4.9 & 2.7 & 8.0 & 2.2 & 4.6 & 4.0   \\
80000 & 10 & 86 & 116 & 111 & 4.2 & 2.5 & 4.1 & 2.2 & 6.2 & 1.7 & 3.5 & 3.5   \\
100000 & 9 & 77 & 104 & 99 & 3.9 & 2.4 & 3.8 & 2.0 & 5.3 & 1.5 & 2.9 & 3.3   \\

\enddata

\tablecomments{The above results are calculated based on a telluric masking treatment in which telluric lines with more than 2\% absorption depth and 30 $\rm{km}\cdot\rm{s}^{-1}$ within its vicinity are masked out. 
}

\end{deluxetable}

\begin{deluxetable}{ccccccccccccc}

%\tabletypesize{\footnotesize}
\tablewidth{400pt}

\tablecaption{Comparison of predicted RV precision (in the unit of $\rm{m}\cdot\rm{s}^{-1}$) between our results to ~\citet{Rodler2011} for an M9.5 dwarf ($T_{\rm{eff}}$=2200 K, $V\sin i$=5 $\rm{km}\cdot\rm{s}^{-1}$)}

\tablehead{\colhead{$\rm{R}$} & \multicolumn{4}{c}{S/N} & \multicolumn{4}{c}{This study} & \multicolumn{4}{c}{~\citet{Rodler2011}} \\
\colhead{} & \colhead{$Y$} & \colhead{$J$} & \colhead{$H$} & \colhead{$K$} & \colhead{$Y$} & \colhead{$J$} & \colhead{$H$} & \colhead{$K$} & \colhead{$Y$} & \colhead{$J$} & \colhead{$H$} & \colhead{$K$} }

%% All data must appear between the \startdata and \enddata commands
\startdata 
\label{tab:Comp_Rodler}
& & & & & & & & & & & & \\
\hline
\multicolumn{13}{l}{Case A} \\
\hline
20000 & 139 & 180 & 171 & 152 & 16.4 & 12.2 & 8.7 & 5.1 & 22.2 & 25.5 & 22.8 & 27.9  \\
40000 & 98 & 127 & 121 & 108 & 9.4 & 6.9 & 5.4 & 3.2 & 6.9 & 8.7 & 7.8 & 10.8  \\
60000 & 80 & 104 & 99 & 88 & 7.7 & 5.5 & 4.6 & 2.8 & 4.2 & 5.7 & 5.1 & 3.8  \\
80000 & 70 & 90 & 85 & 76 & 6.8 & 5.1 & 4.4 & 2.6 & 3.3 & 4.0 & 3.8 & 5.1  \\
\hline
 \multicolumn{13}{l}{Case B} \\
\hline
20000 & 139 & 180 & 171 & 152 & 18.8 & 19.8 & 13.9 & 16.4 & 24.2 & 29.7 & 29.1 & 39.3  \\
40000 & 98 & 127 & 121 & 108 & 10.5 & 10.9 & 8.5 & 9.5 & 8.7 & 12.2 & 11.9 & 17.3  \\
60000 & 80 & 104 & 99 & 88 & 8.6 & 8.7 & 7.0 & 7.8 & 5.4 & 7.1 & 6.8 & 10.7  \\
80000 & 70 & 90 & 85 & 76 & 7.6 & 8.0 & 6.7 & 7.2 & 3.8 & 5.2 & 5.1 & 7.9  \\

\enddata

\tablecomments{Case A is for complete and perfect removal of telluric contamination; Case B is for the case in which telluric lines with absorption depth of $\ge$3\% were masked out. 
}

\end{deluxetable}

\begin{deluxetable}{ccccccccccccc}

\tabletypesize{\scriptsize}
\tablewidth{450pt}

\tablecaption{Required S/N for detection of an Earth-like Planet in the HZ as a function of spectral type}

\tablehead{\colhead{} & \multicolumn{7}{c}{Spectral Band} & \multicolumn{5}{c}{HZ properties} \\
\colhead{} & \colhead{$B$}  & \colhead{$V$} & \colhead{$R$} & \colhead{$Y$} & \colhead{$J$} & \colhead{$H$} & \colhead{$K$} & \colhead{m} & \colhead{$a_{in}$} & \colhead{$a_{out}$} & \colhead{$v_{in}$} & \colhead{$v_{out}$}  \\  \colhead{} & \colhead{} & \colhead{} & \colhead{} & \colhead{} & \colhead{} & \colhead{} & \colhead{} & \colhead{($M_\odot$)} & \colhead{(AU)} & \colhead{(AU)} & \colhead{($\rm{m}\cdot\rm{s}^{-1}$)} & \colhead{($\rm{m}\cdot\rm{s}^{-1}$)}   }

%% All data must appear between the \startdata and \enddata commands
\startdata 
\label{tab:SNR_SP}
A5V & 3483 & 6087 & 14724 & 20389 & 17827 & 35053 & ... & 1.82 &  2.893 &  4.167 &  0.03 &  0.03 \\
F5V & 837 & 1547 & 3088 & 8042 & 7003 & 6976 & ... & 1.30 &  1.614 &  2.325 &  0.06 &  0.05 \\
G5V & 318 & 625 & 1053 & 3149 & 2660 & 2648 & ... & 0.87 &  0.720 &  1.037 &  0.11 &  0.09 \\
K5V & 175 & 280 & 535 & 1710 & 1570 & 1496 & ... & 0.64 &  0.387 &  0.558 &  0.18 &  0.15 \\
M5V & 47 & 58 & 58 & 328 & 359 & 330 & 193 & 0.19 &  0.068 &  0.098 &  0.78 &  0.65\\
M9V & 17 & 22 & 22 & 118 & 99 & 108 & 58 & 0.10 &  0.034 &  0.049 &  1.51 &  1.26 \\

\enddata

\end{deluxetable}

%\clearpage

\begin{deluxetable}{cccccccc}

\tabletypesize{\footnotesize}
\tablewidth{450pt}

\tablecaption{Two examples of telluric contamination}

\tablehead{\colhead{Spectral Type} & \colhead{Bandpass} & \colhead{$\alpha$} & \colhead{$\delta v_{S,rms}$} & \colhead{$\delta v_{ATM,rms}$} & \colhead{$\delta v_{cal}$} & \colhead{$\delta v_{rms}$} & \colhead{$\delta v_{HZ}$} \\
\colhead{} & \colhead{} & \colhead{} & \colhead{($\rm{m}\cdot\rm{s}^{-1}$)} & \colhead{($\rm{m}\cdot\rm{s}^{-1}$)} & \colhead{($\rm{m}\cdot\rm{s}^{-1}$)} & \colhead{($\rm{m}\cdot\rm{s}^{-1}$)} & \colhead{($\rm{m}\cdot\rm{s}^{-1}$)}    }

%% All data must appear between the \startdata and \enddata commands
\startdata 
\label{tab:atm_example}
\multirow{2}{*}{K5V} & \multirow{2}{*}{V} & 1.0 & \multirow{2}{*}{0.11}  & 0.19 & \multirow{2}{*}{0.08}  & 0.24 & \multirow{2}{*}{0.18}  \\
 & & 0.001 & & 0.01 & & 0.14 & \\
 \hline 
\multirow{2}{*}{M5V} & \multirow{2}{*}{K} & 1.0 & \multirow{2}{*}{0.33}  & 7.97 & \multirow{2}{*}{0.14}  & 7.98 & \multirow{2}{*}{0.78}  \\
 & & 0.001 & & 0.17 & & 0.39 & \\

\enddata

\end{deluxetable}

\begin{deluxetable}{ccccc}

\tabletypesize{\footnotesize}
\tablewidth{300pt}

\tablecaption{Prediction vs. HARPS observation}

\tablehead{\colhead{} & \colhead{Bandpass} & \colhead{$\delta v_{S,rms}$} & \colhead{$\delta v_{cal}$} & \colhead{$\delta v_{rms}$}  \\
\colhead{} & \colhead{} & \colhead{($\rm{m}\cdot\rm{s}^{-1}$)} & \colhead{($\rm{m}\cdot\rm{s}^{-1}$)} & \colhead{($\rm{m}\cdot\rm{s}^{-1}$)}     }

%% All data must appear between the \startdata and \enddata commands
\startdata 
\label{tab:harps_example}
$\rm{HD 47186}^a$ & {B+V+R} & ... & {...}  & 0.30  \\
\hline
\multirow{4}{*}{$\rm{G5V}^b$} & {B} & 0.14 & {0.33}  & 0.36  \\
 & {V} & 0.29 & {0.36}  & 0.46  \\
 & {R} & 0.48 & {0.36}  & 0.60 \\
 & {B+V+R} & 0.12 & {0.20}  & 0.24  \\

\enddata
\tablecomments{a: HARPS observation of HD 47186 from~\citet{Bouchy2009}, the best achievable RV precision is at a S/N of 250 for this G5V star with $V\sin i$ of 2.2 $\rm{km}\cdot\rm{s}^{-1}$; b: our RV uncertainties prediction for this star assuming the same S/N, spectral type, observation bandpasses and stellar rotation.
}

\end{deluxetable}

\end{document}